\documentclass[aps,twocolumn,secnumroman, nobalancelastpage,amsmath,amssymb,nofootinbib]{revtex4}
\usepackage{amssymb}
\usepackage{graphicx}
\usepackage{bm}
\usepackage{placeins}
\usepackage{adjustbox}
\usepackage{epstopdf}
\usepackage{morefloats}
\usepackage{blindtext}
\usepackage{array}
\usepackage{makecell}
\usepackage{isotope}

\begin{document}

\title{Systematics of semi-microscopic proton-nucleus optical potential at low energies relevant to nuclear astrophysics}

\author{E. Vagena, M. Axiotis, P. Dimitriou\footnote{Current address: International Atomic Energy Agency, Wagramerstrasse 5, A-1400 Vienna, Austria}}
\email[Corresponding author:]{P.Dimitriou@iaea.org}
\affiliation{Institute of Nuclear and Particle Physics, NCSR ``Demokritos'', 
153.10 Aghia Paraskevi, Athens, Greece}
\date{\today}

\begin{abstract}
\textbf{Background:} Astrophysical models studying the origin of the neutron-deficient p nuclides require knowledge of the reaction rates of neutron, proton and $\alpha$-proton photodisintegrations of pre-existing neutron-rich s- and r-nuclei and of proton capture reaction rates.
Since experimental data at astrophysically relevant interaction energies are limited, reaction rate calculations rely on the predictions of the Hauser-Feshbach (HF) theory. The HF theory requires nuclear physics input such as masses, level densities, $\gamma$-ray strength functions and proton-nucleus optical potentials (OMP) describing the average interaction between the p and the nucleus. The proton OMP plays an important role in the description of proton photodisintegrations and radiative capture reactions at low energies relevant to the p-process nucleosynthesis.

\textbf{Purpose:}
The scope of this work is to improve a global semi-microscopic optical potential for protons at low energies relevant to the p-process nucleosynthesis. This is achieved by adjusting the normalization parameters of the OMP to all available radiative proton-capture cross sections measured at energies of astrophysical interest. By establishing the systematic behaviour of these parameters, one expects to enhance the predictive power of the proton OMP when expanding to mass regions where no data exists.

\textbf{Method:}
The Hauser-Feshbach calculations were  obtained  using the TALYS nuclear reaction code. The normalization parameters for the real and imaginary central potentials ($\lambda_V$ and $\lambda_W$) were adjusted to fit the proton data in the energy range where the cross-section calculations are independent of other input parameters, i.e. neutron optical potential, nuclear level density and $\gamma$-ray strength function. As a consequence, the optimization of the proton OMP was done at energies below the opening of the (\textit{p, n}) reaction threshold. The goodness of the fit is based on the chi-square method as well as on visual comparisons.

\textbf{Results:}
The results show that the normalization parameter $\lambda_V$ of the real part of the proton OMP has a strong mass dependence that can be described by a second degree polynomial function for A $\leq $100 (low mass range) and an exponential increase for $100<$ A $<162$ (intermediate mass range). Though variations of the normalization parameter of the imaginary part $\lambda_W$ have a smaller effect on the calculations, a global increase by 50$\%$ improves the results for certain nuclei without affecting the rest of the cases.

\textbf{Conclusions:}
The resulting adjustment functions were obtained by fitting all suitable proton cross-section data at low energies and can be used with reasonable confidence to generate the global semi-microscopic proton optical potential for nuclei in the medium to heavy mass region. For better statistics, more low-energy proton-capture cross section data are needed for heavier nuclei with mass A $>$ 100.
\end{abstract}

\maketitle

\section{\label{sec:level1} Introduction}

The p process of nucleosynthesis is responsible for the production of the 35 neutron-deficient isotopes -called p nuclei- located along the neutron-deficient side of the chart of nuclides between \isotope[74]{Se} and \isotope[196]{Hg}. Although the exact site for the development of this process is still under investigation, it is generally accepted that it can take place in the oxygen neon layers of massive stars during the type-II supernova (SN) explosion. In such an environment, p nuclei are produced through a complex sequence of neutron, proton and alpha-p photodisintegrations triggered by the heating of pre-existing s- and r-nuclei seeds at temperatures in the range of 1.5 GK and 3.5 GK. Depending on the temperatures and proton density of the surrounding layers, the series of photodisintegrations can be accompanied by proton capture reactions.
Another equally plausible site for the p-process nucleosynthesis is the type-Ia SN explosion resulting from the disruption of the carbon-oxygen (C-O) white dwarf (WD) member of a binary star which has reached a mass close to the Chandrasekhar limit ($M_{Ch}=1.4\times M_{solar}$. Whether one assumes a delayed detonation model of a $M_{Ch}$ WD or a deflagration model of a sub-$M_{Ch}$ WD for the type-Ia SN explosion, the p-process nucleosynthesis can take place in microscopically thin layers heated at temperatures in the range of 2 to 3 GK. The initial composition of the heavy seeds in these layers strongly influences the p-process abundances and is considered to be a major source of uncertainty in these models. However, the general pattern of p-nuclei abundances observed is quite similar to what is observed in massive type-II SN explosions, with the p-nuclides $^{92,94}$Mo, $^{96,98}$Ru, $^{113}$In, $^{115}$Sn, $^{138}$La being underproduced.

Calculations of the abundances of the p nuclei depend on the solution of an extended network of reactions involving about 2000 nuclei in the mass range 12 $\le$ A $\le$ 210 and over 20000 reactions~\cite{arnould2003,rauscher2013}. For protons, in particular, the astrophysically relevant energy range (Gamow window) corresponding to p-process temperatures between 1.8 GK and 3.3 GK is E$_p$ = 1-5 MeV. 
Several sensitivity studies have been performed for p-process nucleosynthesis calculations, identifying a series of ($p,\gamma$), ($\alpha,\gamma$) reactions (and their inverse reactions) that affect the photo-disintegration branchings at certain temperatures~\cite{rauscher2006} or impact the type-II SN p-process abundances~\cite{rapp2006, rauscher2016}.  

Proton-induced reactions also play an important role in the rapid proton (rp)-capture process which occurs in accreting binary systems where one star is a neutron star. The accretion of material onto the neutron star can lead to rising temperatures and eventually to a runaway thermonuclear explosion creating the right conditions, namely temperatures of 1.1 GK to 1.3 GK (E$_p$ = 0.8-2 MeV), for the rp process to occur and produce many neutron-deficient nuclei up to mass A = 100. 
However, it should be noted that this process does not contribute to the p-nuclei abundances since the material produced cannot leave the surface of the neutron star. On the other hand, light p-nuclei can be produced in the proton-rich neutrino wind of type-II SN ($\nu$p process). In this scenario, p nuclei can form at distances where a substantial anti-neutrino flux is present. The latter flux favours the production of heavier species through $(n,p$) reactions followed by a series of neutron and proton captures. The nucleosynthesis in this process is very sensitive to the exact conditions of the neutrino wind, i.e. to the entropy $Y_e$ and the radius~\cite{arnould2003} and is affected by uncertainties given that the exact site of this process has not been established yet. The impact of nuclear uncertainties in a wide range of astrophysical $\nu$p-model conditions has been studied in detail in~\cite{nishimura2019}. 

Due to the large number of reactions involved in the reaction networks relevant to the p-process nucleosynthesis, as well as the difficulties associated with measurements of very small cross sections at low energies close or below the Coulomb barrier, almost all the reactions rates are calculated with the statistical model of Hauser and Feshbach (HF) \cite{hf52}. The nuclear ingredients entering the HF calculations are the nucleon-nucleus optical model potentials (N-OMP), the $\alpha$-p-nucleus optical model ($\alpha$-OMP), the nuclear level densities (NLDs) and the $\gamma$-ray strength functions ($\gamma$SFs). 

Numerous OMPs have been developed to describe the elastic scattering observables of nucleons scattered by nuclei.  These can be classified as (i) local which means that they have been determined for a given nucleus based on the experimental data available for that nucleus at a certain energy or in a range of energies, (ii) regional which means that they apply to nuclei within a small range of masses A at a specific energy or range of energies, and (iii) global, meaning that the OMP has been determined for a wide range of masses A at a given energy range, and can be applied globally to all nuclei at that energy range. The latter may be less
precise than the former two, however they are very practical for nucleosynthesis calculations that involve thousands of nuclei. In the past decades, two global N-OMPs have been developed and used widely to describe nuclear reactions relevant to nuclear astrophysics, the phenomenological N-OMP of Koning and Delaroche~\cite{Koning03} and the semi-microscopic N-OMP of Bauge, Delaroche and Girod~\cite{bauge2001} (JLM/B). For charged-proton reactions in particular, these global OMPs need to be able to describe scattering or reaction observables at low energies associated with the Gamow window for temperatures relevant to the p process (E$_p$ = 1-5 MeV as mentioned above). 

The semi-microscopic JLM/B OMP \cite{bauge2001} has been adjusted to an extensive database of experimental cross sections available at energies ranging from 1 keV to 200 MeV through the introduction of normalization constants ($\lambda_{V,W}$). Although these normalization constants have been thoroughly tested against all available neutron reaction data available at very low energies, for proton-induced reactions they have only been tested against experimental data at energies above 10 MeV. At energies below 10 MeV, the confidence on the adjustable normalization parameters of the proton JLM/B OMP is less than 50$\%$ according to Ref.~\cite{bauge2001}.

The purpose of this work is to explore the applicability of the JLM/B optical potential for proton-induced reactions
at low energies relevant to the astrophysical rp and p processes. A sensitivity study of the parameters of the JLM/B model has been carried out by comparing the calculations with the available proton-capture cross section data. The goal is to establish a systematic behaviour of the proton OMP (pOMP) with respect to mass A that can be used to improve the description of the data over an extended mass region as required in p-process calculations.

The paper is organized as follows: In Section II, the normalization parameters of the JLM/B potential are described. In Section III the selection criteria and limitations of the experimental data are discussed. The results of the systematic comparison between calculations and data are presented in Section IV. Section V presents the conclusions drawn from the present study.

\section{JLM/B optical model potential}
The general functional form of the Lane-consistent JLM/B OMP is given by \cite{bauge2001,bauge1998}:

\begin{equation} \label{eq2}
U=\lambda_{V}[V_{0}\pm \lambda_{V1}\alpha V_{1}]+i\lambda_{W}[W_{0}\pm \lambda_{W1} \alpha W_{1}]+U_{SO}
\end{equation}

where V$_{0}$, W$_{0}$ and V$_{1}$, W$_{1}$ are the real and imaginary isoscalar and isovector components of the central potential seen by a neutron(proton) and U$_{SO}$ is the component due to spin-orbit interaction. $\lambda_{V,W}$ and $\lambda_{V1,W1}$ are the normalization parameters for the real, imaginary, real isovector and imaginary isovector components, that were introduced to adjust the OMP to experimental data. 
In their papers \cite{bauge2001,bauge1998}, Bauge et al. mention that in the energy range of “maximum confidence”, i.e., between 20 and 50 MeV, the uncertainties in $\lambda_{V,W}$ and $\lambda_{V1,W1}$ do not exceed 10$\%$. Outside this energy region, however, larger uncertainties are expected due to the limited or scarce experimental data that were considered at those energies in their fitting procedure. More specifically, they only considered proton-induced reaction data at energies E$_{c.m.}$ $\geq$ 10 MeV in the fitting procedure. For lower energies, $\lambda_{V,W}$, $\lambda_{V1,W1}$ were assumed to be constant and were extrapolated from the aforementioned fits at E$_{c.m.}$ $\geq$ 10 MeV down to E$_{c.m.}$ = 1 keV.

Taking into account the stated uncertainties in the $\lambda_{V,W}$ and $\lambda_{V1,W1}$ normalization parameters, and the fact that low-energy proton-induced data below E$_{c.m.} = 10$ MeV were not included in the determination of these parameters \cite{bauge2001,bauge1998}, we decided to vary the normalization parameters to obtain a better agreement between the  proton-induced experimental cross sections and those calculated using the potential of Bauge et al. \cite{bauge2001,bauge1998} in the low-energy region. We first tested the isoscalar and isovector normalization parameters separately to see the effect on the cross sections. The results show that the isovector $\lambda_{V1,W1}$ have a much weaker effect on the cross sections compared to the isoscalar ones $\lambda_{V,W}$, which is expected since the cross sections depend on the elastic scattering central potential. In our analysis, we therefore adjusted only the isoscalar normalization parameters $\lambda_{V,W}$ to improve the description of the cross section data, while the isovector ones were kept unchanged.

In the following, we introduce the multiplicative factors f$_v$, f$_w$ 
\begin{equation}
    \lambda_V^\prime (E) = \textrm{f}_v \cdot \lambda_V(E) ,\,\,\,\lambda_W^\prime (E) = \textrm{f}_w \cdot \lambda_W(E). \nonumber\label{eq:eq1}
\end{equation}

The factors f$_{v,w}$ correspond to the ``lvadjust" and ``lwadjust" keywords used in the TALYS 1.95 code ~\cite{Talys1.95} to vary the normalization parameters $\lambda_{V,W}$ and range between 0.5 and 1.5 following the prescription of Bauge et al.~\cite{bauge2001}.




\section{Selection of experimental data}\label{sec:exp-data}
In this section the selection of the experimental data used to adjust the new pOMP at low energies is described. We obtained all the (\textit{p,$\gamma$}) cross section data from the EXFOR database \cite{exfor}. A total of 87 (\textit{p,$\gamma$}) cross section datasets were retrieved from EXFOR with the vast majority referring to medium and mid-heavy mass nuclei. 


However, not all of these data were included in the analysis. The selection of the data to be used in the analysis was based mainly on the following requirement: the optimization of the N-OMP is possible only at energies where the HF cross sections depend solely on the N-OMP and are independent of the other ingredients of the HF model, namely the nuclear level density (NLD) and $\gamma$-ray strength function ($\gamma$SF).  In cases where there is sufficient independent experimental information to fix the NLDs and $\gamma$SF models associated with the open reaction channels, then the fitting energy region can be extended to energies where the HF cross section is sensitive to all the mentioned HF nuclear ingredients. However, such cases are rather limited and since our aim was to study the systematics over a wide mass region, we applied the above constraints globally. The main conditions for selecting experimental (p, $\gamma$) cross-section data was that datapoints were available at energies (i) below the opening of the (\textit{p, n}) reaction threshold and (ii) within the energy range where the (p, p$\prime$) and (p, $\alpha$) cross sections are much smaller than the (p, $\gamma$) ones. These conditions end up limiting the suitable experimental data considerably. To be able to cover as wide a mass range as possible, datasets which had at least two datapoints within the above-defined energy range were included in the analysis. Furthermore, data that were published without uncertainties or with limited information on the uncertainty budget were also considered. The available measurements span a period of three decades therefore, they vary in the experimental setups that were used, the precision and accuracy, as well as the provision of information on sources. Due to the limited number of data available in the desired energy region, all the datapoints measured in the desired energy range were included in our analysis, however, where there were issues or doubts about the quality of the data, a smaller weight was assigned to these data in the fitting process. Light nuclei (A $<40$) were not included in the analysis as we only used the statistical HF model for the calculations.

From the 87 available (p, $\gamma$) datasets, only those for 30 nuclei with atomic number (Z) from 22 to 68 and mass number (A) from 47 to 162 were found to be suitable for the analysis. The nuclei considered in the analysis are listed in Table~\ref{tab:experimentalData}. The experimental data that is proposed to be re-measured due to the limited number of datapoints at the energy region of interest or due to discrepancies between datasets are marked with an asterisk. Most of the target nuclei that are used in this analysis have mass A $<100$. For mass A $>100$ there are fewer suitable data available, while for A $>162$ there are no experimental proton capture data at low energies due to the experimental challenges involved in measuring very small cross sections at energies near the Coulomb barrier. The energy ranges considered in the analysis are listed in Table~\ref{tab:experimentalData}. The determination of the energy ranges is discussed in Sects.~\ref{sec:strength}, ~\ref{sec:ldensity}. 

\begin{table}[b]
\caption{\label{tab:experimentalData}%
The proton capture reactions included in the analysis, the neutron-emission threshold S$_n$ (\textit{p,n}) reaction, number of datapoints and the upper energy limit E$_{max}$ taken into account in the fitting procedure. Asterisk (*) marks all the proposed data to be measured again either due to the limited number of datapoints at the energy region of interest or due to the discrepancies found between datasets and / or between datasets and theory.
}
\begin{ruledtabular}
\begin{tabular}{c||c||c||c||c}
\textrm{Nucleus}&
\textrm{\makecell{S$_n$\\(MeV)}} &
\textrm{\makecell{E$_{max}$\\(MeV)}}&
\textrm{\makecell{Datapoints}} &
\textrm{References} \\
\colrule
\isotope[47]{Ti}$^*$ & 3.71 & 0.9 & 4 & \cite{kennett81}\\ 
\isotope[48]{Ti}$^*$ & 4.80 & 0.9 & 2 & \cite{kennett81}\\
\isotope[49]{Ti}$^*$ & 1.38 & 0.9 & 8 & \cite{fedorets1986,kennett80}\\
\isotope[51]{V}$^*$ & 1.53  & 1.0 & 2 & \cite{zyskind80}\\
\isotope[53]{Cr} & 1.38 & 1.4 & 14 &
\cite{gardner81}\\
\isotope[54]{Cr} & 2.16 & 1.4 & 11 &
\cite{zyskind78}\\
\isotope[58]{Fe} & 3.09 & 1.1 & 4 &
\cite{tims93}\\
\isotope[59]{Co}$^*$ & 6.91 & 1.2 & 4 &
\cite{butler57}\\
\isotope[60]{Ni} & 6.91 & 1.2 & 3 &
\cite{tingwell89}\\
\isotope[61]{Ni} & 3.02 & 1.3 & 9 &
\cite{tingwell88}\\
\isotope[65]{Cu} & 2.13 & 1.5 & 10 &
\cite{sevior83}\\
\isotope[74]{Ge} & 3.34 & 2.0 & 3 &
\cite{quinn2013}\\
\isotope[77]{Se}$^*$ & 2.15 & 2.0 & 3 &
\cite{krivonosov1977}\\
\isotope[86]{Sr} & 6.02 & 2.0 & 6 &
\cite{qyurky2001}\\
\isotope[87]{Sr} & 2.64 & 2.6 & 8 &
\cite{qyurky2001}\\
\isotope[88]{Sr} & 4.40 & 2.8 & 15 &
\cite{galanopoulos2003}\\
\isotope[89]{Y} & 3.61 & 2.3 & 10 &
\cite{harissopulos13,tsagari04} \\
\isotope[92]{Zr} & 2.79 & 2.8 & 5 &
\cite{spyrou2013}\\
\isotope[92]{Mo} & 8.66 & 1.8 & 3 &
\cite{sauter97}\\
\isotope[94]{Mo} & 5.04 & 2.5 & 11 &
\cite{sauter97}\\
\isotope[96]{Mo} & 3.76 & 2.8 & 7 &
\cite{foteinou2019}\\
\isotope[98]{Mo} & 2.45 & 2.5 & 4 &
\cite{foteinou2019}\\
\isotope[98]{Ru}$^*$ & 5.83 & 2.8 & 15 &
\cite{bork1998}\\
\isotope[104]{Pd} & 5.06 & 3.2 & 5 &
\cite{dillmann2011,spyrou08}\\
\isotope[106]{Pd} & 3.75 & 3.7 & 6 &
\cite{spyrou08}\\
\isotope[108]{Cd} & 5.91 & 3.0 & 5 &
\cite{gyurky07}\\
\isotope[120]{Te} & 6.40 & 3.2 & 2 &
\cite{guray09}\\
\isotope[130]{Ba}$^*$ & 6.42 & 3.8 & 2 &
\cite{netterdon2014}\\
\isotope[152]{Gd} & 4.77 & 4.8 & 3 &
\cite{guray2015}\\
\isotope[162]{Er} & 5.64 & 5.6 & 4 &
\cite{ozkan2017}\\

\end{tabular}
\end{ruledtabular}
\end{table}

\section{Model calculations and results}
As already mentioned, nucleon-induced reactions on medium- and/or medium-heavy mass nuclei at energies relevant to the rp and p process take place mainly through the compound-nucleus reaction mechanism described by the Hauser-Feshbach (HF) statistical model~\cite{hf52}. The calculated $\sigma_{HF}$ depends on the choice of the models for the main ingredients of the $\sigma_{HF}$, namely the nucleon-nucleus optical model potential (N-OMP), the (NLD) and the $\gamma$-ray strength function ($\gamma$SF). 
Since our aim is to improve the pOMP of JLM/B by comparing HF calculations with existing experimental data, the first step in the analysis is to determine the energy range where the $\sigma_{HF}$ does not depend on the neutron OMP, level density and $\gamma$-ray strength but only on the pOMP. 

In this work, the cross sections calculations were performed with the latest version (1.95) of the nuclear reaction code TALYS \cite{Talys1.95}.

\subsection{\label{sec:strength}Strength-function models}

 The sensitivity of the calculated $\sigma_{HF}$ to the $\gamma$SF function was investigated by looking at how the TALYS cross sections varied with the various $\gamma$SF models implemented in the code.
 Calculations were performed using different $\gamma$SF models while keeping the OMP and NLD
 models unchanged. The level density was set to the Constant Temperature-Fermi gas model (CTFG) ~\cite{cftg-talys} which is the default option of the TALYS code while the nucleon OMP was set to the JLM/B~\cite{bauge2001} option. 
 The $\gamma$SF models that were used include the Kopecky-Uhl generalized Lorentzian (KU)~\cite{ku1990}, Brink-Axel Lorentzian (BA)~\cite{brink1957,axel1962}, Hartree-Fock BCS tables (HFBCS/QRPA)~\cite{goriely2002}, Hartree-Fock-Bogolyubov tables (HFB/QRPA)~\cite{goriely2004}, Goriely’s hybrid model (HG)~\cite{goriely1998}, T-dependent HFB (HFB/T)~\cite{goriely2004}, T-dependent RMF (RMF/T)~\cite{daoutidis2012} and the Gogny D1M HFB+QRPA (DIM/HFB/QRPA)~\cite{martini2016,goriely2019}.
  The impact of the different $\gamma$SF functions on the calculated (\textit{p, $\gamma$}) cross sections for the cases listed in Table~\ref{tab:experimentalData} is shown in Fig.~\ref{fig:fig1}. The Gamow window for the rp and p process corresponding to temperature ranges of (1.1 - 1.3 GK) and (1.8 - 3.3 GK), respectively, is also indicated in each plot. 
 
 The results show that the different $\gamma$SFs have a significant impact on the cross sections within the p-process Gamow window, which in some cases can vary by a 2-3 orders of magnitude, for almost all the studied cases. This confirms that the $\sigma_{HF}$ is sensitive to the $\gamma$SF over a large part of the Gamow window.
On the other hand, there is a limited energy range in the lower end of the Gamow window, where the cross sections  are insensitive to the $\gamma$SF models. This energy range increases with increasing mass of the target nucleus. This is the energy range which we shall use to adjust the pOMP in this work. In this limited energy range the HF cross sections depend on the OMP in the incident channel only, i.e. the pOMP whereas the neutron OMP only plays a role above the (p, n) reaction threshold. 

Consequently, by focusing on the description of proton-induced reactions at low energies below the neutron threshold, we are testing and improving the pOMP only. This of course implies that the Lane Consistency of the original JLM/B potential is not preserved, as the resulting normalization parameters $\lambda_{V,W}$ of the potential will be different for protons and neutrons.

\subsection{\label{sec:ldensity}Level density models}

The sensitivity of the calculations to the NLDs associated with the open reaction channels was explored in a similar fashion as in the case of the $\gamma$SF. 
Various calculations were performed using the different phenomenological and microscopic level density models available in TALYS 1.95, i.e. Constant temperature Fermi gas (CTFG) which is specific to TALYS~\cite{cftg-talys,koning2008}, Back-shifted Fermi gas (BSFG)~\cite{dilg1973,koning2008}, Generalised superfluid model (GSM)~\cite{ignatyuk1979,ignatyuk1993}, Hartree-Fock-BCS (HFBCS)~\cite{demetriou2001}, Hartree-Fock-Bogolyubov (HFB)~\cite{goriely2008} and Temperature-dependent
Hartree-Fock-Bogolyubov (HFB/T)~\cite{hilaire2013}. 
The $\gamma$SF was set to the Kopecky-Uhl generalized Lorentzian model~\cite{ku1990} which is the default option in the TALYS code while the nucleon OMP was set to the JLM/B~\cite{bauge2001} model. The results obtained with the different NLD models are plotted in Fig.~\ref{fig:fig2}. From the figure it is clear that the different NLD models have a minor impact on the $\sigma_{HF}$ within the Gamow energy window compared to the $\gamma$SFs. As a result, the energy range that will be used to determine the pOMP can be determined from the sensitivity of the $\gamma$SF function solely.

\begin{figure*}[ht]
\centering
\includegraphics[trim={0 0cm 0 0cm},clip,width=\textwidth]{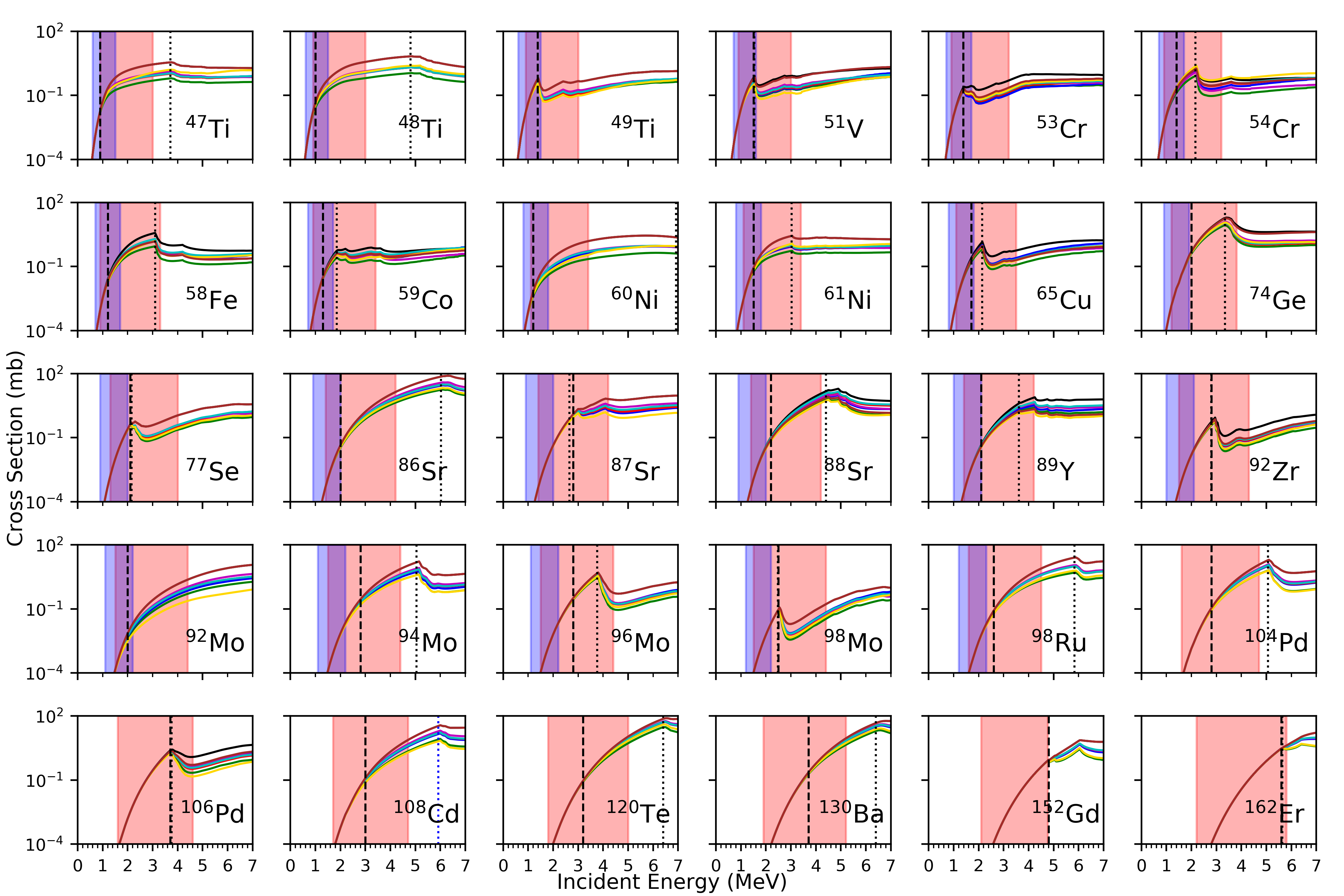}
\caption{Comparison of calculated (\textit{p, $\gamma$}) cross sections using all possible $\gamma$SF functions available in TALYS (i.e. Kopecky-Uhl (green), Brink-Axel (black), HFBCS/QRPA (blue),  HFB/QRPA (red), Hybrid-Goriely (magenta), HFB/T (cyan), RMF/T (yellow) and DIM/HFB/QRPA (brown)). The dashed lines show the upper energy limit below which the calculations are independent of the $\gamma$SF. The shaded areas indicate the Gamow window for the rp (blue color) and p process (red color). The end-point of the rp process is around A = 100. The dotted lines indicate the neutron-emission threshold S$_n$.}\label{fig:fig1}
\end{figure*}

\begin{figure*}[ht]
\centering
\includegraphics[trim={0 0cm 0 0cm},clip,width=\textwidth]{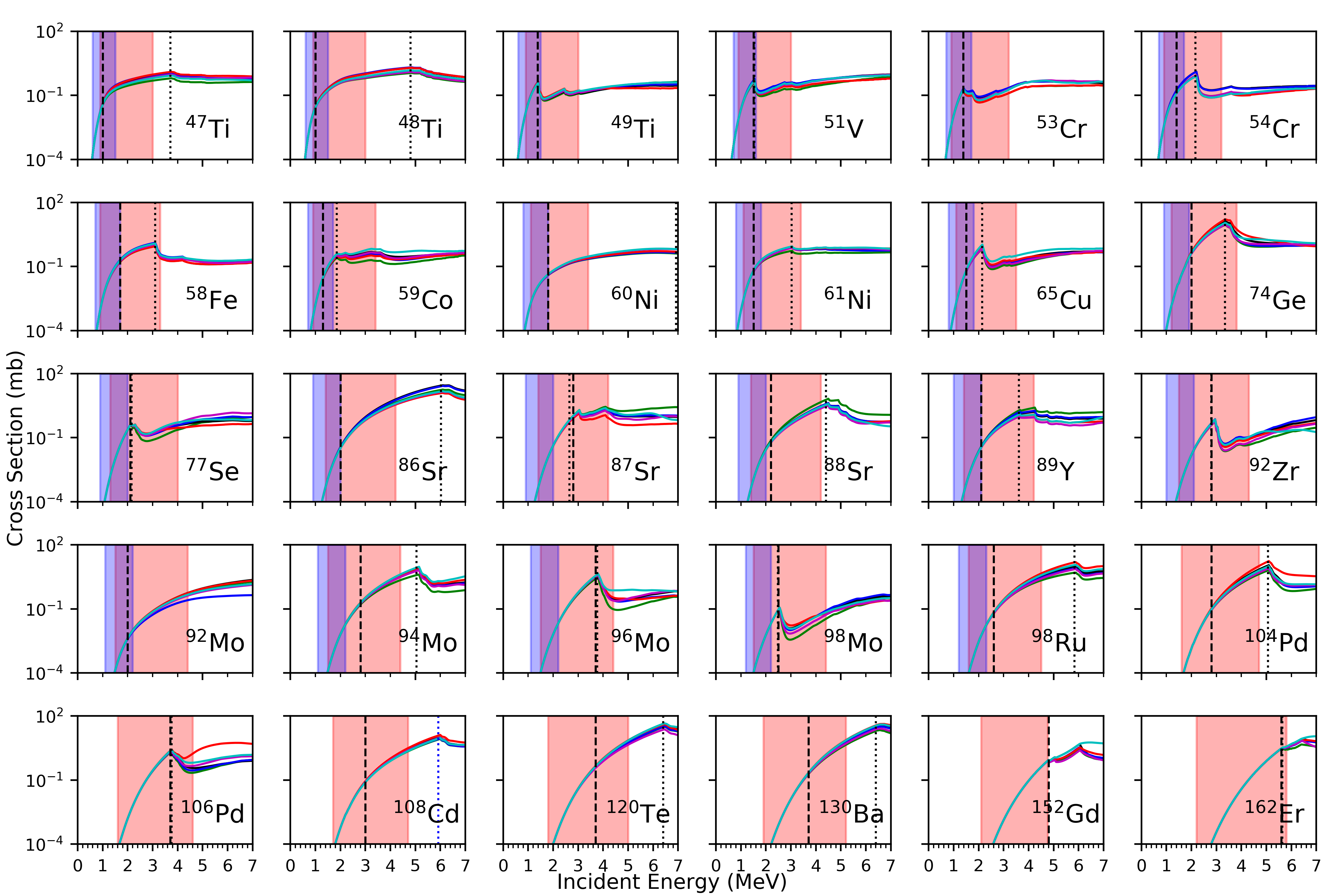}
\caption{Comparison of calculated (\textit{p, $\gamma$}) cross sections using all possible NLD models available in TALYS (i.e. CTFG (green), BSF (black), GSM (blue),  HFBCS (red), HFB (magenta), HFB/T (cyan)). The dashed lines show the upper energy limit below which the calculations are independent of the different NLD models. The shaded areas indicate the Gamow window for the rp (blue color) and p process (red color). The dotted lines indicate the neutron-emission threshold S$_n$.}\label{fig:fig2}
\end{figure*}

Using the results of the sensitivity studies shown in this section and in Sect.~\ref{sec:strength}, and bearing in mind that the neutron OMP plays a role only after the opening of the (\textit{p, n}) channel (see Figs.~\ref{fig:fig1}-\ref{fig:fig2}), we have determined the energy ranges in which the cross sections are exclusively sensitive to the pOMP (displayed in Table~\ref{tab:experimentalData}).

\subsection{\label{sec:level2}Parameter search}

The JLM/B pOMP was adjusted for each nucleus separately by searching for the values of both $\lambda_{V,W}$ normalization parameters that reproduce the proton-capture cross sections in the energy ranges listed in Table~\ref{tab:experimentalData}. The energy dependence of $\lambda_{V,W}$ was not  modified but was kept the same as in Ref.~\cite{bauge2001}. The $\lambda_{V,W}$ parameters were varied by applying the multiplicative factors f$_{v,w}$ as shown in Eq.~\ref{eq:eq1}. The search for the multiplicative factors f$_{v,w}$ was performed in two stages. 

In the first stage, calculations were performed for each nucleus by varying the f$_v$ factor between 0.5 and 1.5 using a step of 0.1. A smaller step in the variation of f$_v$ was also tested (i.e. 0.01 and 0.001) only to reveal the non-linear relation between the $\sigma_{HF}$ and the f$_v$ factor. The non-linear relation between $\sigma_{HF}$ and the f$_v$ implies that there is no simple and unique trend function that describes the best fit values for all the listed nuclei. In these calculations the $\lambda_W$ factor was kept unchanged at its default value (1.0). 

\begin{figure}[!h] 
  \centering
  \includegraphics[trim={0.cm 0.cm 0.cm 1cm},clip,width=.5\textwidth]{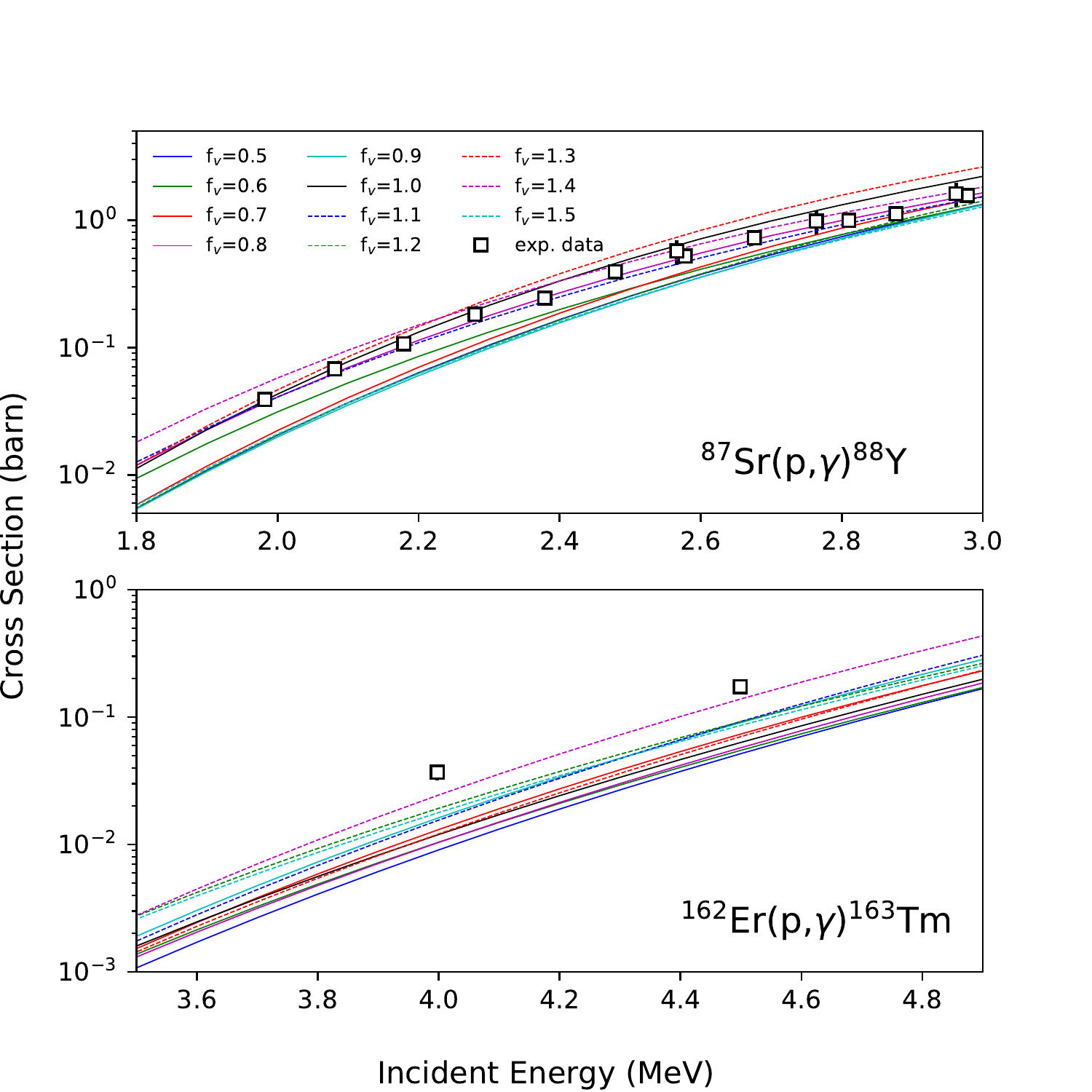}
  \caption{The (\textit{p, $\gamma$}) experimental cross sections for \isotope[87]{Sr} (top) and \isotope[162]{Er} (bottom) of Ref. \cite{qyurky2001} and \cite{ozkan2017}, respectively, compared to model calculations using different values of f$_v$ for the JLM/B pOMP. For \isotope[87]{Sr}, more than one value of f$_v$ factor describes the experimental data well while for \isotope[162]{Er} only f$_v = 1.4$ can reproduce the data.}  
  \label{fig:fig3}
\end{figure}

It should be noted that the goodness of fit was determined using both the chi-square method and visual comparison. 
This was necessary because of the non-uniform nature of the experimental uncertainties of the available experimental data. In some cases, experimental uncertainties were missing altogether, while in other cases it was not clear whether the assigned uncertainties were purely statistical or included systematic errors as well. As a result, the chi-square analysis was applied without consideration of experimental errors (weighting factors), and visual comparison was used to ensure that additional weight was placed on the fit in energy regions where the data were expected to be more reliable. 
The non-linear relation between $\sigma_{HF}$ and f$_v$ meant that the fitting procedure yielded more than one best fit values for the f$_v$ factor. For example, for the \isotope[87]{Sr}(\textit{p,$\gamma$}) reaction, a variation of f$_v$ by both 20$\%$ (f$_v=0.8$) and 10$\%$ (f$_v=1.1$) results in an equally good description of the experimental data as shown in Fig.~\ref{fig:fig3}. However, in certain cases such as the \isotope[162]{Er}(\textit{p,$\gamma$}) reaction, also shown in Fig.~\ref{fig:fig3}, we found a unique value of the multiplicative factor f$_v$ that can reproduce the data fairly well, namely f$_v  = 1.4$ (a 40$\%$ increase). To conclude, in some cases the final best value of the normalization factor was selected from a set of best fit values according to how well it matched the global trend of all the best fit values with respect to the nuclear mass A. 

\begin{figure}[!h] 
  \centering
  \includegraphics[trim={0.5cm 0.5cm 2.cm 2cm},clip,width=.4\textwidth]{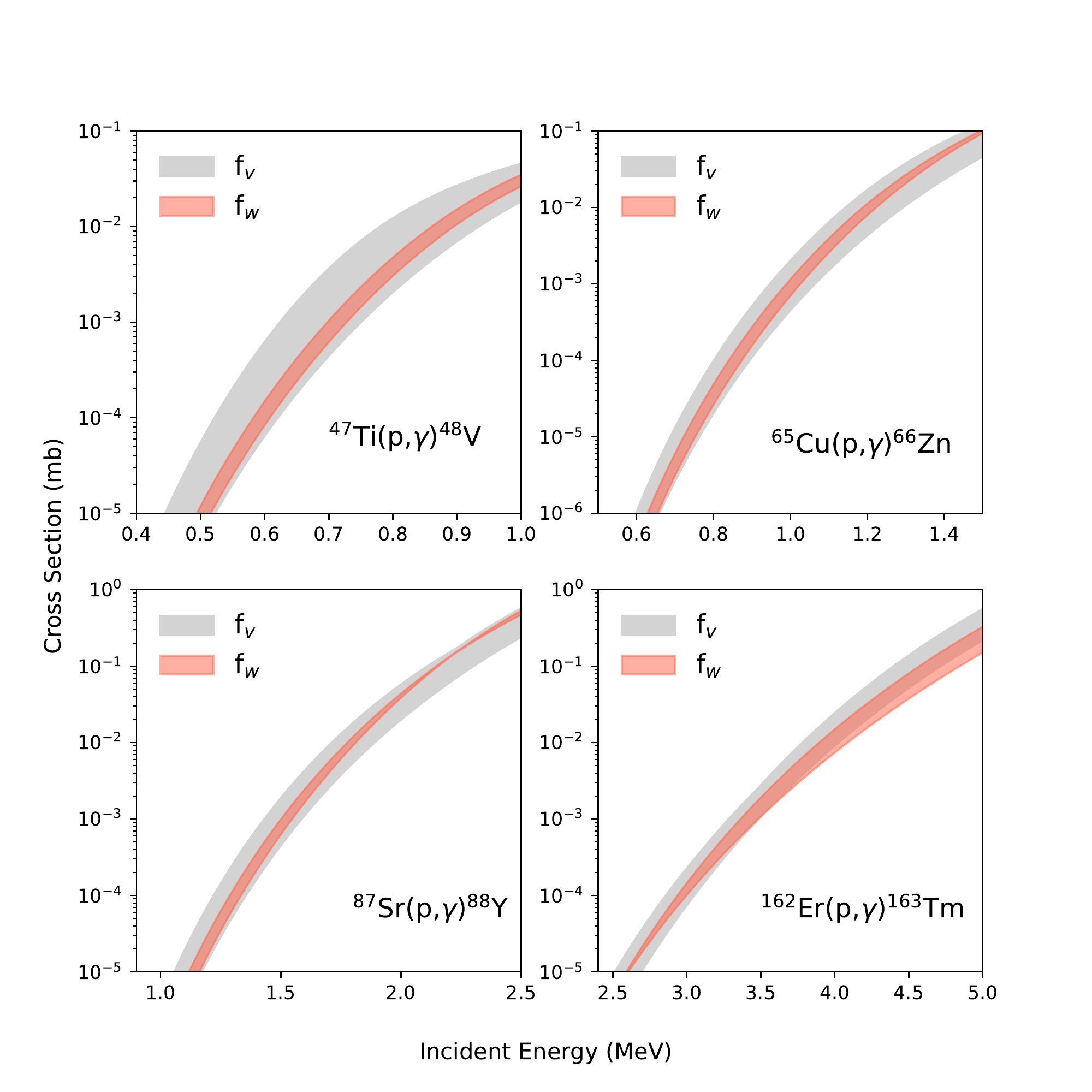}
  \caption{Ranges of calculated cross sections obtained by varying the f$_v$ (grey shade) and f$_w$ (orange shade) factors separately, for nuclides \isotope[47]{Ti}, \isotope[65]{Cu}, \isotope[87]{Sr}, \isotope[162]{Er}. In all cases the imaginary part of the JLM/B pOMP (f$_w$) has a smaller impact on the cross sections in the fitted energy range. }
  \label{fig:fig4}
\end{figure}

In the second stage, the results obtained in the 1st stage were further improved by using the best fit values of f$_v$ and adjusting the f$_w$ factor between 0.5 and 1.5 with a step of 0.1. From the comparison with the data it is clear that the f$_w$ factor, which influences the imaginary part of the JLM/B pOMP, has a smaller impact on the cross sections within the fitting energy range. In Fig.~\ref{fig:fig4}, we compare the results obtained for all possible values of the f$_v$ and f$_w$ factors for four cases representing a broad mass range, namely \isotope[47]{Ti}, \isotope[65]{Cu}, \isotope[87]{Sr}, \isotope[162]{Er}.

\FloatBarrier

The results of the two-stage fitting process for all the thirty nuclei included in Table~\ref{tab:experimentalData} show that there is a trend in the values of the f$_v$ and f$_w$ factors with respect to the nuclear mass A. Specifically, for A $\leq$ 100 the f$_v$ values decrease with increasing A, while for A $>100$ they increase with increasing A. As was mentioned above, for cases where the data could be reproduced fairly well with more than one f$_v$ values (e.g.\isotope[87]{Sr}), the final value was chosen based on the general trend. This trend is valid and verified for nuclei with mass up to A = 162 (\isotope[162]{Er}) as listed in Table~\ref{tab:experimentalData}.

\begin{figure}[!htbp] 
  \centering
  \includegraphics[trim={0.5cm 0.cm 1.cm 1cm},clip,width=.5\textwidth]{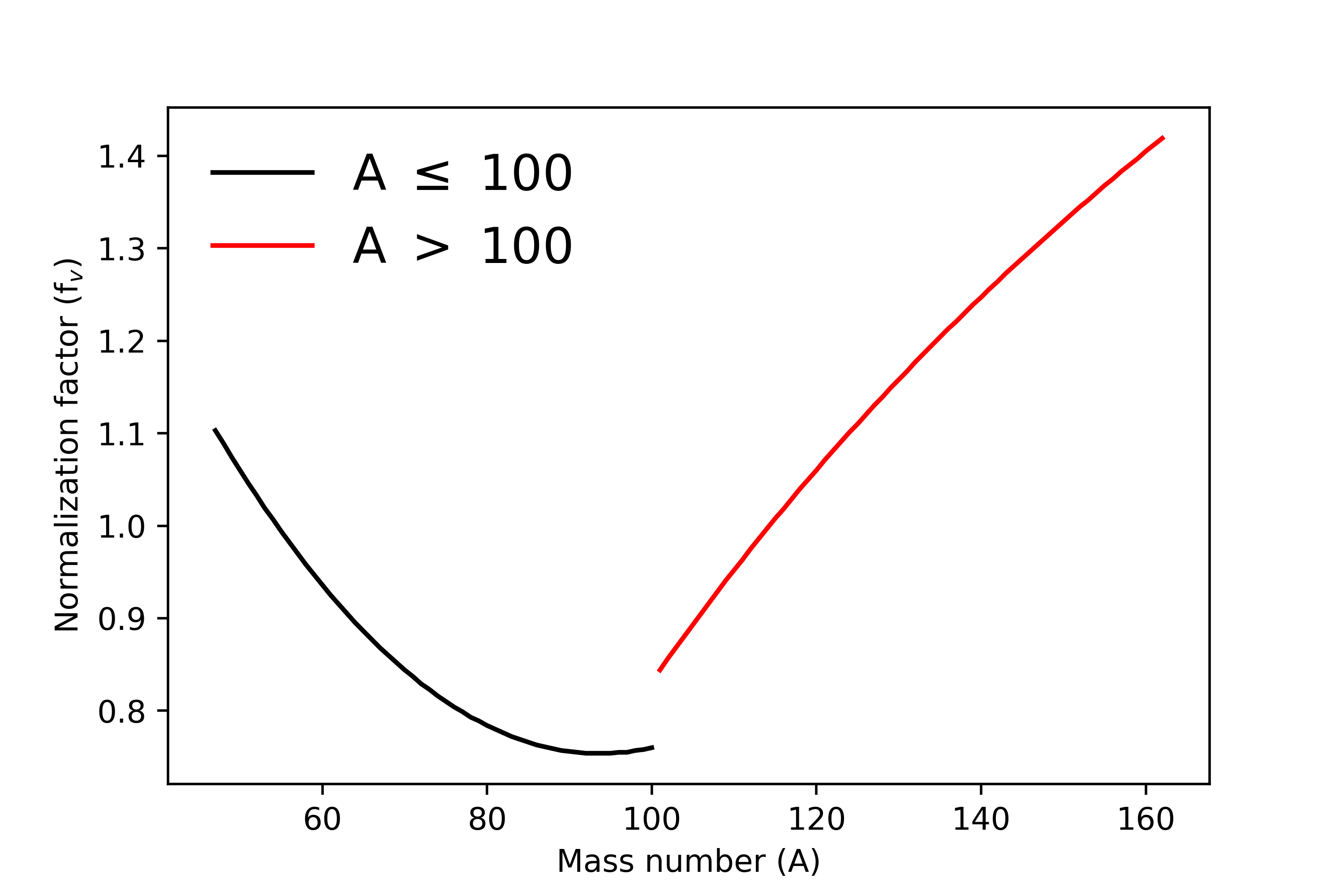}
  \caption{Mass dependence of the correction factor f$_v$ of the real part of the pOMP for mass regions A $\leq$ 100 and A $>100$.}
  \label{fig:fig5}
\end{figure}

The A-dependence of f$_v$ can be described by a second degree polynomial function for A $\leq$ 100 and by a logarithmic increase for A $>100$, as follows (also shown in Fig.~\ref{fig:fig5}): 

For A $\leq$ 100:
\begin{equation}
{\textrm{f}{_v}=0.00016 A^2 - 0.03 A + 2.16}  \hspace{8mm}%
\nonumber
\end{equation}

For A $>100$:
\begin{equation}
{\textrm{f}{_v}=5.9 \ln{(0.25 \ln{A})}}  \hspace{8mm}%
\label{eq:eq3}
\end{equation}

The final values of f$_v$ obtained from the above functions are summarized in Table~\ref{tab:table2}. For A $\leq$ 100, the normalization factor f$_v$ of the real part of the JLM/B pOMP decreases smoothly with increasing nuclear mass and for nuclei around A $\sim$ 100 this decrease can be by as much as 20$\%$. On the other hand, for A $>100$, the factor f$_v$ increases steeply with mass A and reaches the maximum value (1.4) for \isotope[162]{Er}. 
The agreement between data and calculations can be further improved by increasing the normalization factor f$_w$ of the imaginary part of the pOMP by 50$\%$ (f$_w=1.5$) for all the nuclei in Table~\ref{tab:experimentalData}.

\begin{table}[h!]
  \begin{ruledtabular}
    \caption{Normalization factors for the real central pOMP obtained from the fitting process. Values are rounded to the first digit.}
    \label{tab:table2}
    \begin{tabular}{lc||lc||lc||lc} 
     Nucleus & f$_v$ & Nucleus & f$_v$ & Nucleus & f$_v$ & Nucleus & f$_v$\\
      \hline
      \isotope[47]{Ti} & 1.1  & \isotope[60]{Ni} & 1.0 &  \isotope[89]{Y} & 0.8  &\isotope[108]{Cd} & 0.9\\
      \isotope[48]{Ti} & 1.1  & \isotope[61]{Ni} & 0.9 &  \isotope[92]{Zr}  & 0.8  &\isotope[120]{Te} & 1.1\\
      \isotope[49]{Ti} & 1.1  & \isotope[65]{Cu} & 0.9 &  \isotope[94]{Mo} & 0.8  &\isotope[130]{Ba} & 1.2\\
      \isotope[51]{V}  & 1.1  & \isotope[74]{Ge} & 0.8 &  \isotope[96]{Mo} & 0.8  &\isotope[152]{Gd} & 1.3\\
      \isotope[53]{Cr} & 1.0  & \isotope[77]{Se} & 0.8 &  \isotope[98]{Mo} & 0.8  &\isotope[162]{Er} & 1.4\\
      \isotope[54]{Cr} & 1.0  & \isotope[86]{Sr} & 0.8 &  \isotope[98]{Ru} & 0.8  &\\
      \isotope[58]{Fe} & 1.0  & \isotope[87]{Sr} & 0.8 &  \isotope[104]{Pd} & 0.9  &\\
      \isotope[59]{Co} & 1.0  & \isotope[88]{Sr} & 0.8 &  \isotope[106]{Pd} & 0.9  &\\
    \end{tabular}
    \end{ruledtabular}
\end{table}

\begin{figure*}[!hp]
    \centering
    \includegraphics[trim={2cm 4cm 2cm 4cm},clip,width=\textwidth]{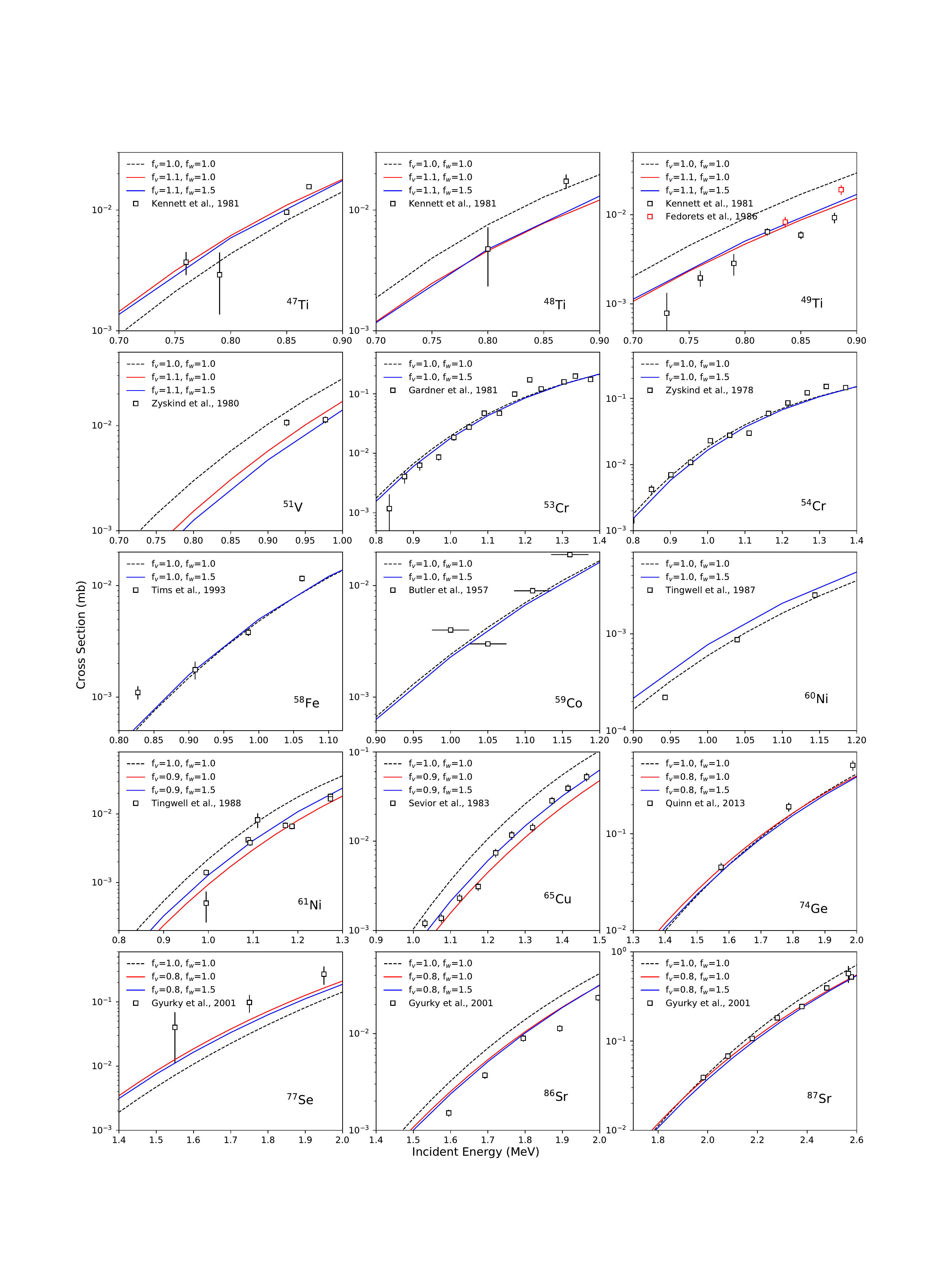}
    \caption{Comparison between the experimental (\textit{p, $\gamma$}) cross sections and the calculated cross sections obtained with (i) standard JLM/B potential (dotted black line), (ii) adjusted JLM/B pOMP using f$_{v}$ from Eq.~\ref{eq:eq3} and default value f$_{w}=1.0$ (red line) and (iii) adjusted JLM/B pOMP using f$_{v}$ from Eq.~\ref{eq:eq3} and f$_{w}=1.5$ (blue line). }
    \label{fig:fig6}
\end{figure*}

\begin{figure*}[!hp]
    \centering
    \includegraphics[trim={2cm 4cm 2cm 4cm},clip,width=\textwidth]{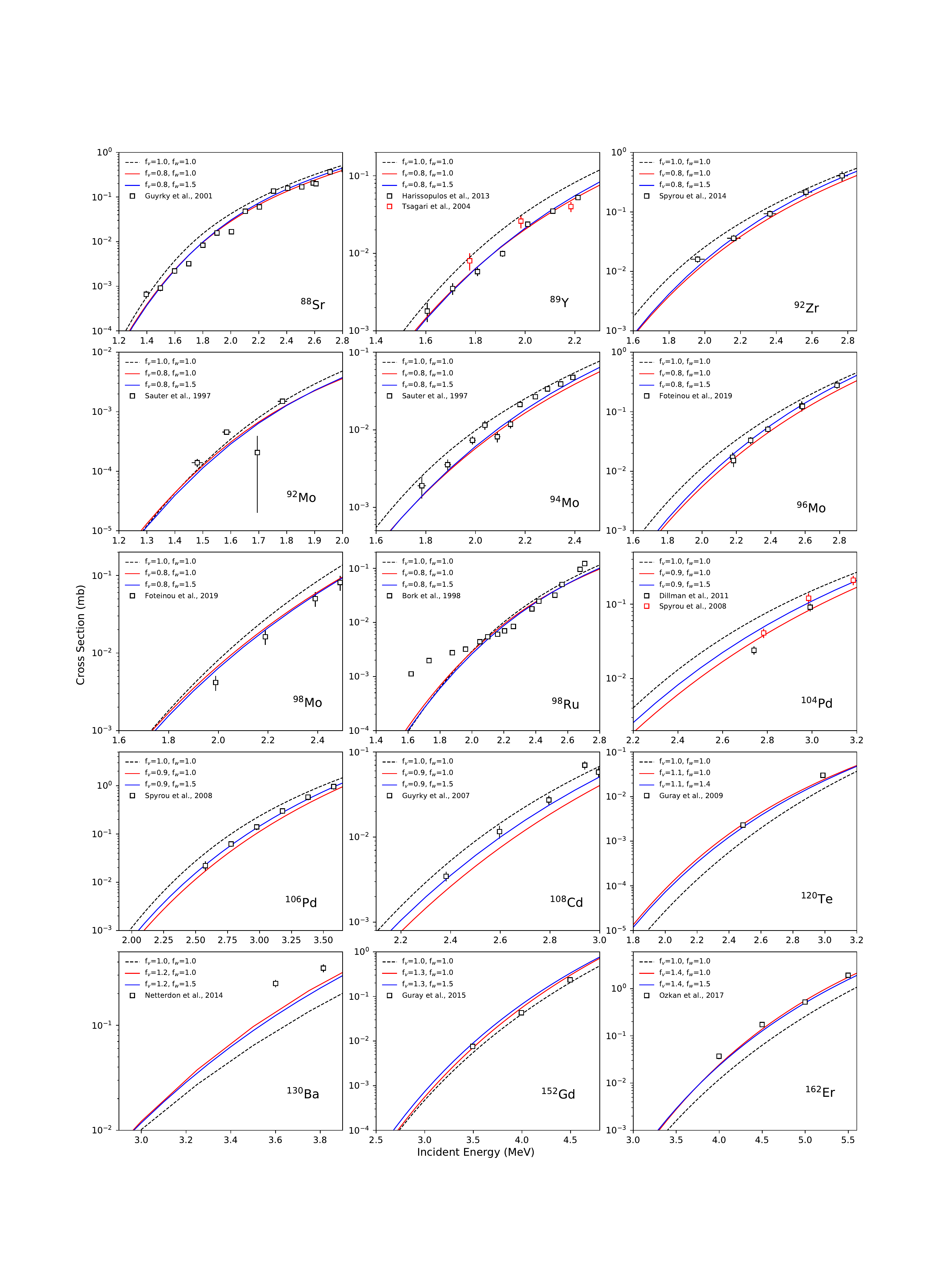}
    \caption{Same as in Fig.~\ref{fig:fig6}.}
    \label{fig:fig7}
\end{figure*}

In Figs.~\ref{fig:fig6}-\ref{fig:fig7} we compare the experimental (\textit{p, $\gamma$}) cross sections with the calculated cross sections obtained using the standard JLM/B of ~\cite{bauge2001} (f$_{v,w} = 1$) (dotted black line) and the adjusted JLM/B pOMP with f$_{v}$ from Eq.~\ref{eq:eq3} combined with f$_{w}=1.0$ (red line) and f$_{w}=1.5$ (blue line), respectively. Overall, the new adjustments factors f$_{v,w}$ lead to an improved agreement between experiment and theory. The suggested f$_{v,w}$ values reproduce the experimental data fairly well, except for a few isolated cases. For A $\leq$ 100 in particular, the adjustment factors f$_{v,w}$ improve the agreement with the data significantly compared to the default values (f$_{v,w}=1$) which lead to deviations from the data by a factor of 2 at least. In a few cases (i.e. \isotope[87,88]{Sr}), the default and suggested f$_v$ values yield comparable results. Overall, one observes that for light and mid-heavy nuclei, f$_v=1$ overestimates the experimental cross sections, whereas for heavier nuclei such as, \isotope[120]{Te},\isotope[130]{Ba},\isotope[152]{Gd},\isotope[162]{Er}, it underestimates the data. The increase in the adjustment factor f$_w$ by 50$\%$ improves the agreement with the experimental data for certain nuclei such as \isotope[106]{Pd} and \isotope[108]{Cd}.

\section{Impact of improved proton optical potential on (\textit{p, n}) reactions}\label{sec:pn-data}
The impact of the modified pOMP on the (p, n) cross sections at low energies within the Gamow window has been explored for nuclei with mass A = 49-181 for which experimental data are available. At these low incident energies, the dominant reaction mechanism in (p, n) reactions is the compound nucleus mechanism which is described by the HF statistical model as detailed in the previous sections. Just as in the case of (p, $\gamma$) reactions, the relevant nuclear quantities in the calculation of the HF cross sections are the nucleon OMPs, NLDs and $\gamma$SFs, and in the limited energy region where the HF cross section depends solely on the N-OMP, it is the pOMP we are testing.  The nucleon OMP that is relevant in these calculations is shown in Eq.~\ref{eq:eq1} while contributions from the excitation of isobaric analogue states are not considered in this work. 

The calculations were performed with the TALYS 1.95 nuclear reaction code as described in previous sections. The (p, n) cross sections obtained with the same nuclear input as for the (p, $\gamma$) cross sections were compared with experimental data for 51 nuclei listed in Table~\ref{tab:table3}. The (p, n) data were retrieved from EXFOR and were checked for necessary corrections in the centre-of-mass energy due to energy losses in the target. The impact of the other nuclear ingredients of the HF calculations, such as neutron OMP, NLDs and $\gamma$SF were also investigated. For the purpose of validating the pOMP, we only considered those cases where the influence of these other nuclear properties on the (p, n) cross sections is limited to a small energy range of about 200 keV around the reaction threshold and for which experimental data existed above this small range. All other cases were excluded as they would not allow us to draw any conclusion on the pOMP.

Furthermore, there are also issues with some of the measurements dating back to the early 50s and 60s. For example, the data of Blaser et al.~\cite{blaser1951} for \isotope[87]{Rb}, \isotope[87]{Sr}, \isotope[96]{Zr}, \isotope[107]{Ag}, \isotope[111]{Cd}, and \isotope[128]{Te} suffer from large corrections at energies above 3.5 MeV due to beam straggling effects that are not reflected in the given experimental uncertainties, while Johnson et al.~\cite{johnson1964} report uncertainties of about a factor of 2. Details of the measurements by Skakun et al.\cite{skakun1987} for \isotope[100]{Mo} are not accessible so we cannot comment on the observed discrepancies.

To summarize, including (p, n) reactions in our analysis allows us to verify and validate the systematics of the pOMP at energies over the entire Gamow window, and for a larger number of nuclei that were not accessible with the (p, $\gamma$) reactions only. Our results however, underscore the need for new additional measurements on (p, n) cross sections at low energies relevant to the p-process Gamow window to resolve the observed discrepancies and allow for an unambiguous validation of the parameters of the pOMP.

\begin{table}[h]
\caption{\label{tab:table3}%
The (p, n) reactions considered in the comparisons. Asterisk (*) marks all the proposed data to be measured again either due to the limited number of datapoints at the energy region of interest or due to the discrepancies found between datasets and / or between datasets and theory.
}
\begin{ruledtabular}
\begin{tabular}{cl||cl}
\textrm{Nucleus}&
\textrm{References}&
\textrm{Nucleus}&
\textrm{References}\\
\colrule
\isotope[49]{Ti} & \cite{johnson1958,johnson1964,kennett1980} &
\isotope[51]{V} &  \cite{johnson1958,johnson1964,harris1965,zyskind80,mehta1977, barrandon1975}\\
\isotope[53]{Cr} & \cite{johnson1958,johnson1964,gardner81} &
\isotope[54]{Cr} & \cite{johnson1960,johnson1964,zyskind78, kailas1975}\\
\isotope[55]{Mn} & \cite{johnson1958,johnson1964,mitchell1983}&
\isotope[57]{Fe} & \cite{johnson1964}\\
\isotope[58]{Fe}$^*$ & \cite{tims93}&
\isotope[59]{Co} & \cite{johnson1958,johnson1964,kailas1975}\\
\isotope[65]{Cu} & \cite{switkowski1978,sevior83,saini1983,hershberger1984,generalov2017}&
\isotope[67]{Zn}$^*$ & \cite{johnson1964,blaser1951}\\
\isotope[71]{Ga} & \cite{johnson1958,johnson1964}&
\isotope[76]{Ge} & \cite{kiss2007}\\
\isotope[75]{As} & \cite{johnson1958,johnson1964,albert1959}&
\isotope[77]{Se} & \cite{johnson1958,johnson1960,johnson1964}\\
\isotope[80]{Se} & \cite{johnson1964,foteinou2018, kailas1979}&
\isotope[82]{Se} & \cite{johnson1958,debuyst1968,foteinou2018, gyurky2003}\\
\isotope[79]{Br} & \cite{west1993}&
\isotope[81]{Br} & \cite{west1993}\\
\isotope[85]{Rb} & \cite{kiss2008}&
\isotope[87]{Rb}$^*$ & \cite{blaser1951}\\
\isotope[87]{Sr} & \cite{blaser1951}&
\isotope[92]{Zr} & \cite{blaser1951,flynn1979}\\
\isotope[94]{Zr} & \cite{flynn1979, fedorets1977}&
\isotope[96]{Zr}$^*$ & \cite{blaser1951}\\
\isotope[93]{Nb} & \cite{johnson1958,johnson1964,albert1959}&
\isotope[95]{Mo} & \cite{flynn1979, skakun1987}\\
\isotope[96]{Mo} & \cite{flynn1979}&
\isotope[98]{Mo} & \cite{flynn1979}\\
\isotope[100]{Mo}$^*$ & \cite{skakun1987}&
\isotope[103]{Rh} & \cite{johnson1960,johnson1964,sudar2002, blaser1951}\\
\isotope[110]{Pd} & \cite{johnson1960,johnson1964}&
\isotope[107]{Ag} & \cite{blaser1951,wing1962,hershberger1980}\\
\isotope[109]{Ag} & \cite{johnson1960,johnson1964,dmitriev1967}&
\isotope[111]{Cd}$^*$ & \cite{blaser1951,skakun1975}\\
\isotope[116]{Cd} & \cite{johnson1958,johnson1964}&
\isotope[115]{In} & \cite{johnson1960,johnson1964,hershberger1980}\\
\isotope[117]{Sn} & \cite{klucharev1970,johnson1970,johnson1977,batij1991,blaser1951}&
\isotope[119]{Sn} & \cite{johnson1970,johnson1977}\\
\isotope[122]{Sn} & \cite{klucharev1970,johnson1970,johnson1977}&
\isotope[124]{Sn} & \cite{johnson1970,johnson1977,elmaghraby2009}\\
\isotope[128]{Te}$^*$ & \cite{blaser1951}&
\isotope[130]{Te}$^*$ & \cite{johnson1958,johnson1964}\\
\isotope[127]{I} &  \cite{colle1974,west1993}&
\isotope[139]{La}$^*$ & \cite{wing1962}\\
\isotope[142]{Ce}$^*$ & \cite{verdieck1967}&
\isotope[147]{Sm} & \cite{gheorghe2014}\\
\isotope[149]{Sm} & \cite{gheorghe2014}&
\isotope[151]{Eu}$^*$ & \cite{west1989}\\
\isotope[153]{Eu}$^*$ & \cite{west1989}&
\isotope[169]{Tm} & \cite{spahn2005,sonnabend2011}\\
\isotope[181]{Ta}$^*$ & \cite{hansen1962}

\end{tabular}
\end{ruledtabular}
\end{table}

\begin{figure*}[!hp]
    \centering
    \includegraphics[trim={2cm 4cm 2cm 4cm},clip,width=\textwidth]{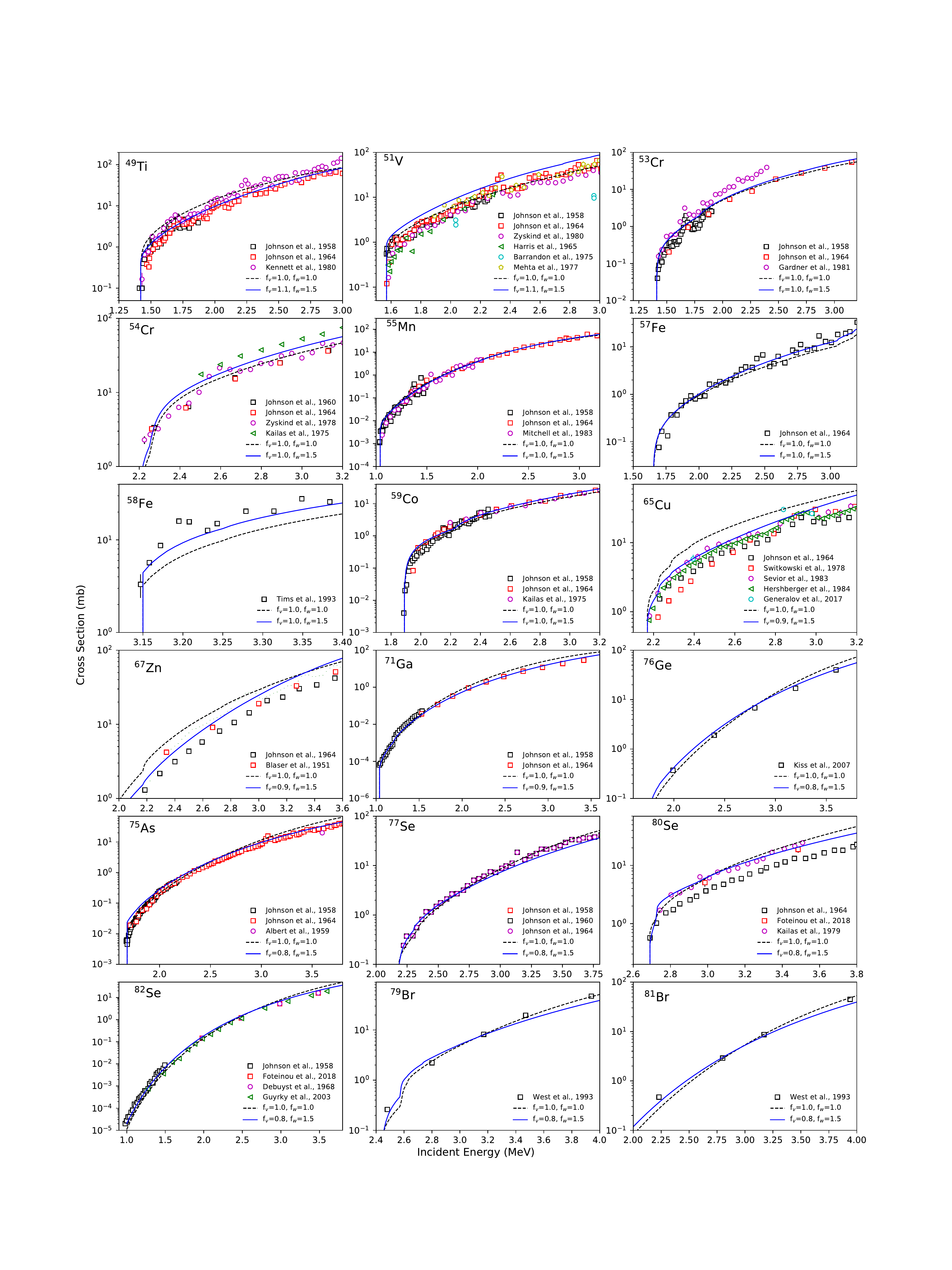}
    \caption{Comparison of measured (p, n) cross sections and TALYS calculations using both the standard (dotted black line) and modified proton JLM/B pOMP obtained using the adjusted values of f$_{v,w}$ (blue line).}
    \label{fig:fig8}
\end{figure*}

\begin{figure*}[!hp]
    \centering
    \includegraphics[trim={2cm 4cm 2cm 4cm},clip,width=\textwidth]{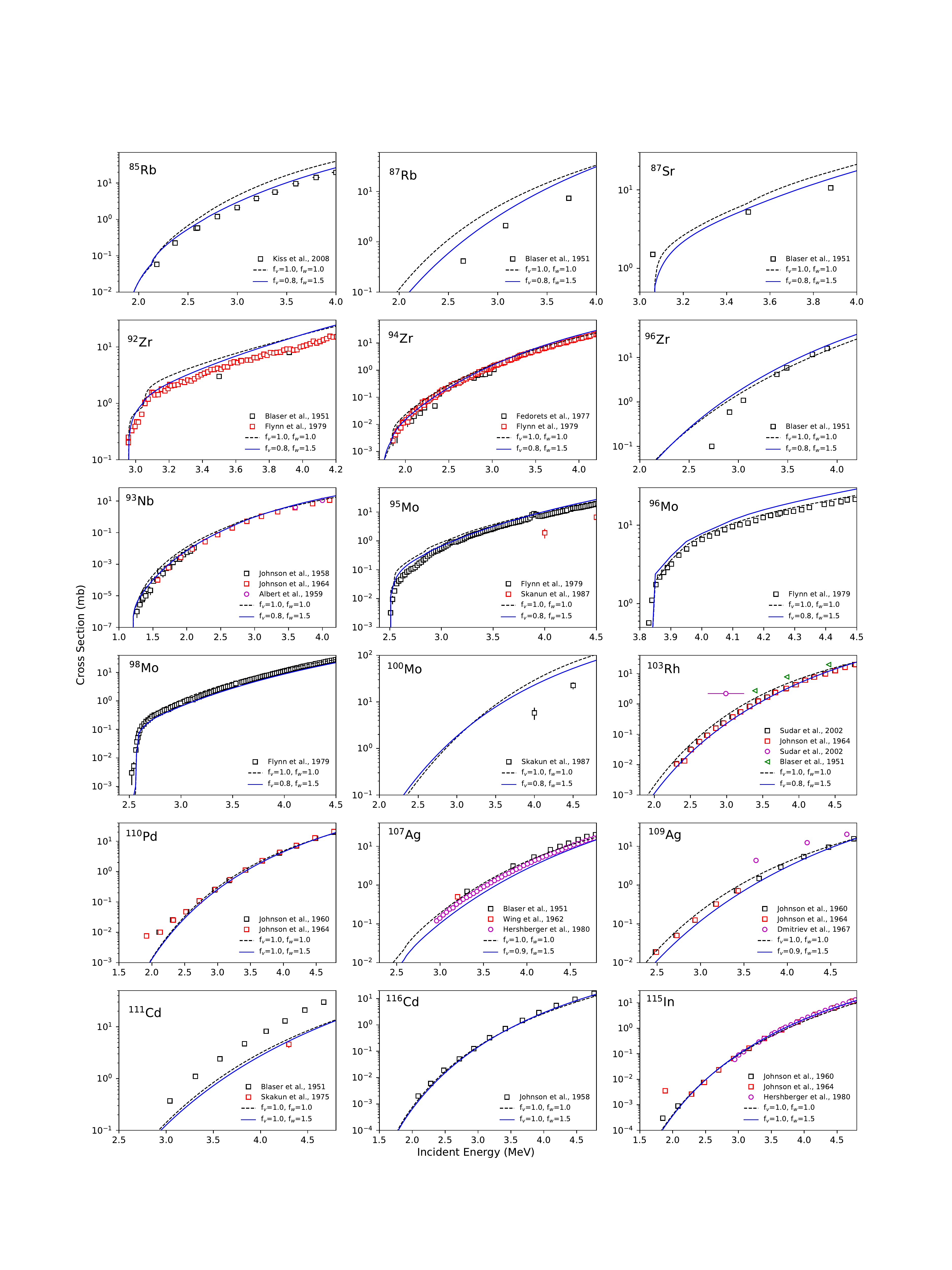}
    \caption{Comparison of measured (p, n) cross sections and TALYS calculations using both the standard (dotted black line) and modified proton JLM/B pOMP obtained using the adjusted values of f$_{v,w}$ (blue line).}
    \label{fig:fig9}
\end{figure*}

\begin{figure*}[!hp]
    \centering
    \includegraphics[trim={2cm 4cm 2cm 4cm},clip,width=\textwidth]{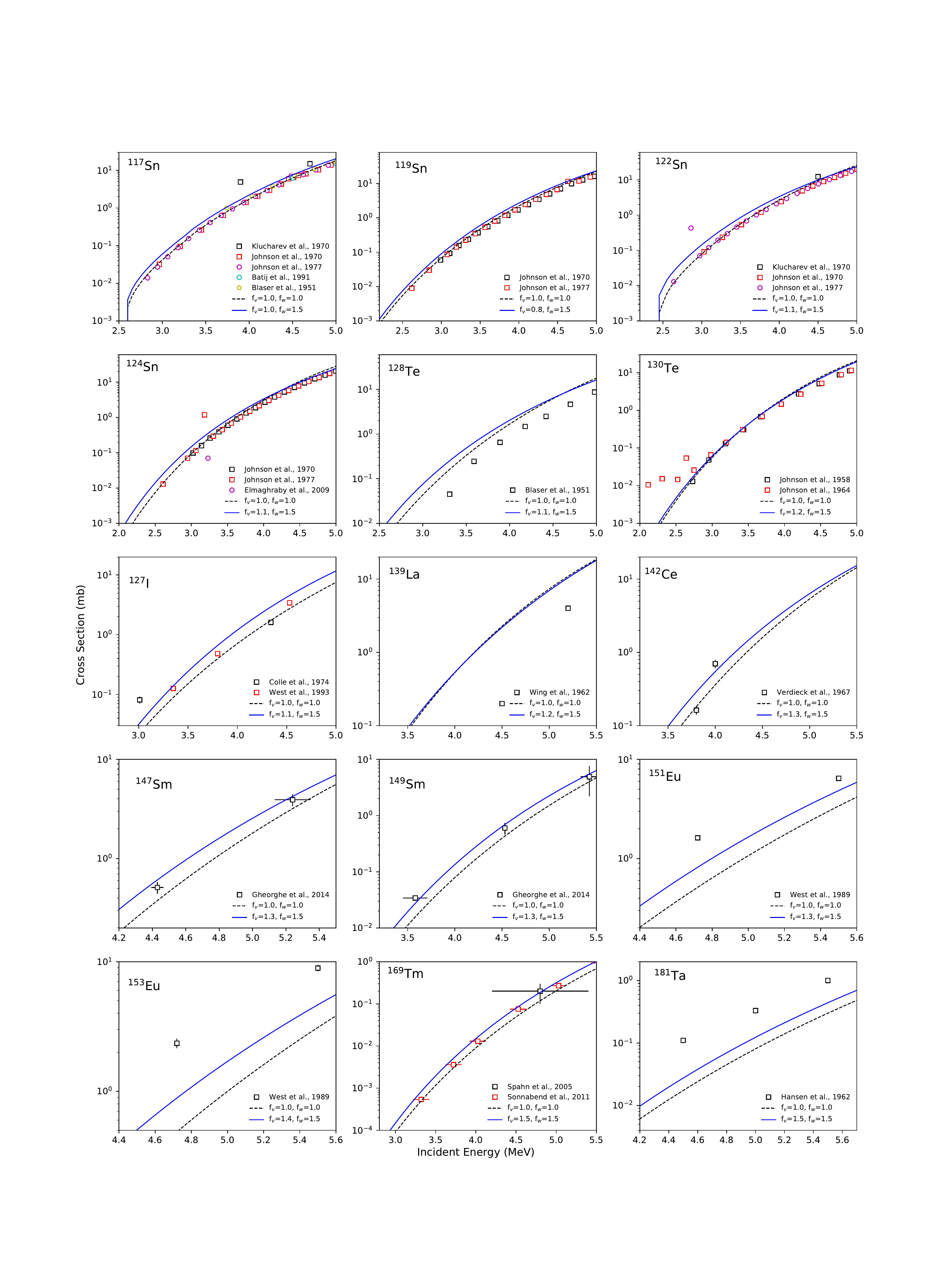}
    \caption{Comparison of measured (p, n) cross sections and TALYS calculations using both the standard (dotted black line) and modified proton JLM/B pOMP obtained using the adjusted values of f$_{v,w}$ (blue line).}
    \label{fig:fig10}
\end{figure*}
\FloatBarrier

\section{Impact of improved proton OMP on large-scale calculations}
To investigate the impact of the new pOMP on a larger set of reactions, involving unstable targets for which no experimental data are available, Maxwellian averaged cross sections (MACS) were calculated for all nuclei with $6 \leq Z \leq 84$ lying between the proton drip line and the valley of stability that are potentially relevant to the p-, vp- and rp-processes of nucleosynthesis. Figures 10 and 11 show the ratio of the MACS obtained with the modified over the default JLM /B model on the (N,Z) plane at the temperatures of 1.5 GK and 3 GK, respectively. As can be seen, the differences in the MACS reach a maximum factor of 3.5 at $T_9 = 1.5$ GK.  On average the differences across the neutron-deficient part of the nuclear chart are of the order of 50\%. The highest deviations by a factor of 3 to 3.5 are found for $\sim$ 50 radioactive nuclei with $50 \leq Z \leq 80$ and $120 \leq A \leq 160$ at the lower temperature of 1.5 GK. Large differences, by a factor of 2 - 2.8, are also obtained for several stable isotopes which are listed in Table~\ref{tab:table4}. The latter could be measured in the lab to allow for a better determination of the correction factors as described in Sect.~\ref{sec:level2}.

These above-mentioned differences observed in the MACS when using the modified and default pOMPs are not as significant as the differences found when using various low-energy $\alpha$-nucleus OMPs~\cite{demetriou2002} confirming what is widely accepted, namely that at low energies the proton OMP is known with better accuracy than the $\alpha$-nucleus OMP. 

\begin{figure*}[!htbp] 
  \centering
  \includegraphics[trim={0.5cm 0.cm 1.cm 1cm},clip,width=\textwidth]{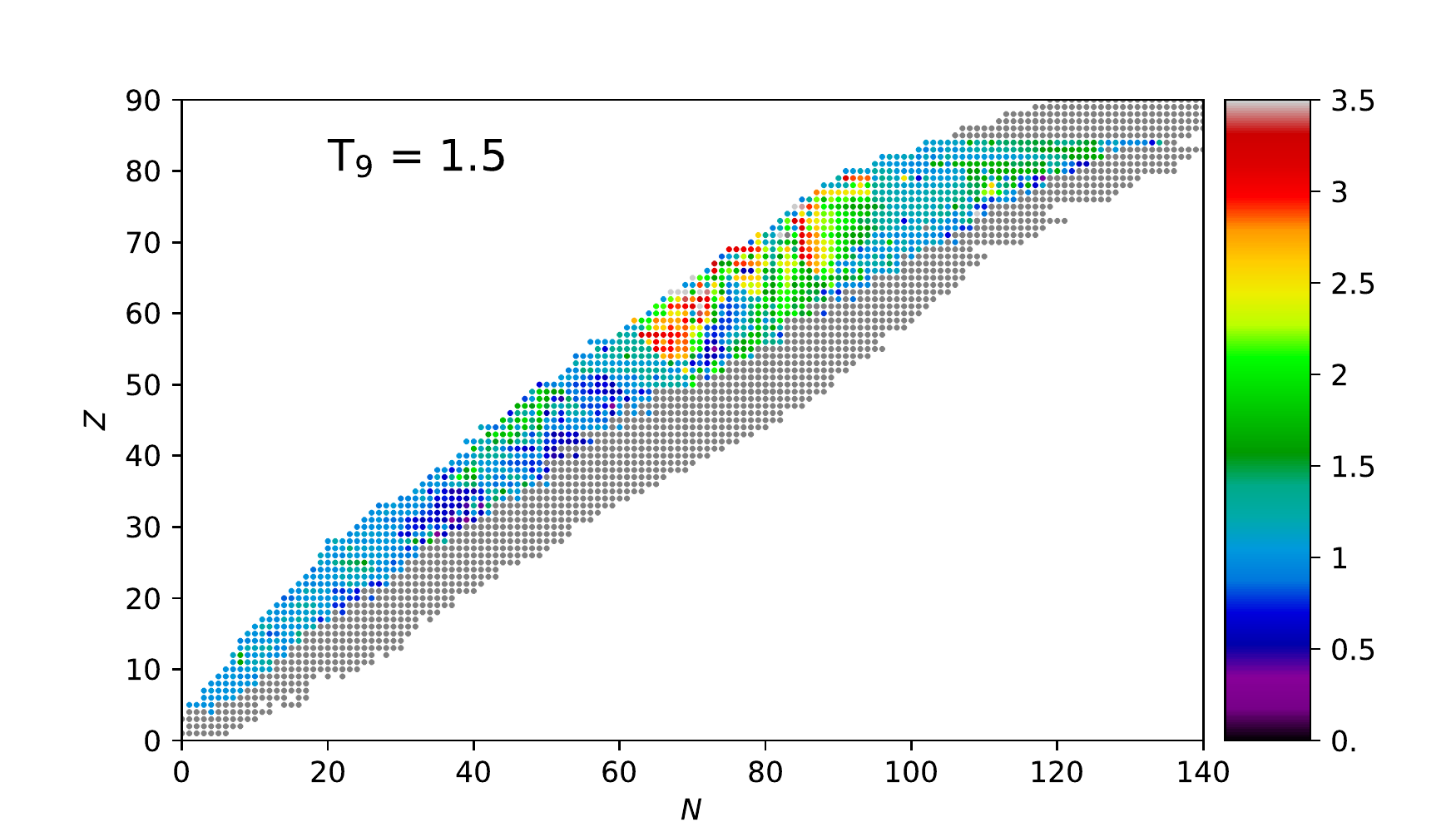}
  \caption{(N,Z)-plane of the MACS ratio between the modified and default JLM model at 1.5 GK.}
  \label{fig:fig11}
\end{figure*}

The impact of the differences observed in Figs.~(\ref{fig:fig10}-\ref{fig:fig11}) on nucleosynthesis calculations remains to be studied. Seeing that the largest differences, by a factor of 3 - 3.5, are obtained at the lower temperatures that are relevant to the SN type-II p-process site, it is expected that p-process abundance calculations in this scenario will be more sensitive to the modified JLM/B pOMP than in the SN type-Ia scenario described in Sect.~\ref{sec:level1}.

Table~\ref{tab:table4} lists the reactions that are proposed for measurement. The recommendations are based on the issues affecting the existing experimental data as detailed in Sects.~\ref{sec:exp-data} and \ref{sec:pn-data}. Also listed are the cases for which the MACS mentioned above differ the most and experimental data are not available. 

\begin{table}[h]
\centering
\caption{\label{tab:table4}%
List of proton-induced reactions recommended for measurement at low energies. Problems with experimental data include i) few data available in the fitting energy range, ii) discrepant data, and iii) data without experimental errors.
}
\begin{ruledtabular}
\begin{tabular}{ c || l  || l}
\textrm{(p,$\gamma$)} &  \isotope[47]{Ti}, \isotope[48]{Ti},  \isotope[49]{Ti},  \isotope[51]{V}, & Problems with exp. data \\
 & \isotope[59]{Co},   \isotope[77]{Se},   \isotope[98]{Ru},   \isotope[130]{Ba} & Figs.~(\ref{fig:fig6}-\ref{fig:fig7})\\
  &  & \\
  
 \textrm{(p,n)} & \isotope[58]{Fe}, \isotope[69]{Ga},  \isotope[100]{Mo}, \isotope[94]{Zr}, & Same as above  \\
 & \isotope[128]{Te}, \isotope[197]{Au}, \isotope[142]{Ce}, \isotope[196]{Pt}, & Figs.~(\ref{fig:fig8}-\ref{fig:fig9}) \\ 
 & \isotope[153]{Eu}, \isotope[139]{La}, \isotope[67]{Zn}, \isotope[87]{Rb}, \isotope[96]{Zr} & \\ 
  & & \\
 
\textrm{(p,$\gamma$)}  & \isotope[126]{Te}, \isotope[122]{Sn}, \isotope[120]{Sn}, \isotope[144]{Sm}, & most affected by pOMP  \\
 & \isotope[142]{Nd}, \isotope[125]{Te}, \isotope[197]{Au} & no exp. data \\
  & & Figs.~(\ref{fig:fig11}-12)\\

\end{tabular}
\end{ruledtabular}
\end{table}

\begin{figure*}[!htbp] 
  \centering
  \includegraphics[trim={0.5cm 0.cm 1.cm 1cm},clip,width=\textwidth]{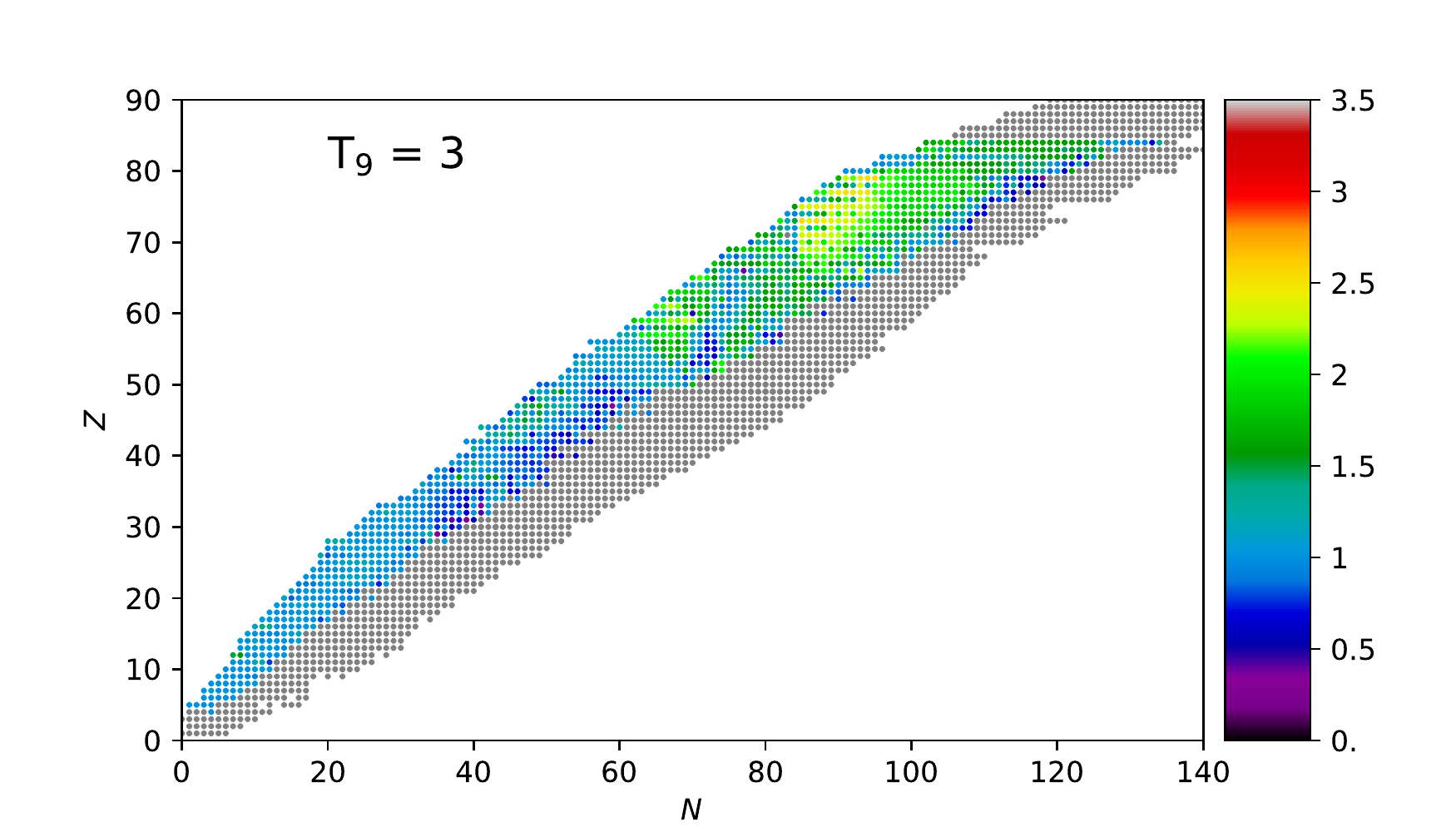}
  \caption{(N,Z)-plane of the MACS ratio between the modified and default JLM model at 3 GK.}
  \label{fig:fig12}
\end{figure*}
\FloatBarrier

\section{Conclusions}
The parameters of a semi-microscopic global proton optical model have been investigated for a wide range of nuclei at low energies relevant to nuclear astrophysics. The $\lambda_{V,W}$ normalization parameters of the real and imaginary components of the semi-microscopic JLM/B potential of Bauge et al.~\cite{bauge2001} have been adjusted to the experimental proton-capture cross section data in the Gamow energy window. 
The results show that the $\lambda_V$ parameter of the real part of the potential has a strong mass dependence which displays two separate trends, a polynomial decrease with increasing mass for A $\leq$ 100 and a logarithmic increase for A $>100$. The imaginary component $\lambda_{W}$ has a smaller effect on the calculations, however we find that an increase by 50$\%$ improves the description of the data for certain nuclei while not affecting the overall majority of cases studied.

The adjusted proton OMP has been validated by systematic comparisons between calculated (p, n) cross sections and available experimental data in the energy range of interest. Overall a good agreement has been found for the majority of cases studied.

It is important to note that the validity of the systematics with respect to mass A is limited by the scarcity of experimental data on proton capture reactions at low energies below the neutron threshold, especially in the heavier mass region (A $>160$). The quality of several of the measured datasets, in particular, the lack of experimental uncertainties or of a detailed and traceable uncertainty budget also affects the accuracy and reliability of the deduced trend functions. The inclusion of (p, n) reaction cross sections in the analysis allows us to extend the energy range beyond the neutron threshold to cover the whole Gamow window relevant to p process temperatures, as well as the upper mass limit from A = 162 to A = 181. However, as was observed in the (p, $\gamma$) measurements, quite often the available experimental data lend themselves to a qualitative rather than a quantitative comparison with calculations. As a result, new measurements of (p,$\gamma$) and (p,n) reactions cross sections are proposed to address these issues and allow for robust and global systematics of the proton OMP at low energies, particularly in the heavier mass region (A $>100$).

The next steps in our effort to improve the proton JLM/B optical potential at low energies relevant to nuclear astrophysics, is a) to investigate the energy dependence of the normalization parameters $\lambda_{V,W}$ in addition to the mass dependence, as we expect that this would improve the agreement between calculations and data in those cases where we observe deviations both in absolute scale and in shape, and b) implement the improved proton OMP p-process calculations using various p-process scenarios to assess the impact of the improved nuclear data on p-nuclei abundances.

\begin{acknowledgements}

{We acknowledge support of this work by the project “CALIBRA/EYIE” (MIS 5002799), which is implemented under the Action “Reinforcement of the Research and Innovation Infrastructures”, funded by the Operational Programme “Competitiveness, Entrepreneurship and Innovation” (NSRF 2014-2020) and co-financed by Greece and the European Union (European Regional Development Fund).}
\end{acknowledgements}

\bibliography{apssamp}

\begin{thebibliography}{100}
\expandafter\ifx\csname natexlab\endcsname\relax\def\natexlab#1{#1}\fi
\expandafter\ifx\csname bibnamefont\endcsname\relax
  \def\bibnamefont#1{#1}\fi
\expandafter\ifx\csname bibfnamefont\endcsname\relax
  \def\bibfnamefont#1{#1}\fi
\expandafter\ifx\csname citenamefont\endcsname\relax
  \def\citenamefont#1{#1}\fi
\expandafter\ifx\csname url\endcsname\relax
  \def\url#1{\texttt{#1}}\fi
\expandafter\ifx\csname urlprefix\endcsname\relax\def\urlprefix{URL }\fi
\providecommand{\bibinfo}[2]{#2}
\providecommand{\eprint}[2][]{\url{#2}}

\bibitem[{\citenamefont{Arnould and Goriely}(2003)}]{arnould2003}
\bibinfo{author}{\bibfnamefont{M.}~\bibnamefont{Arnould}} \bibnamefont{and}
  \bibinfo{author}{\bibfnamefont{S.}~\bibnamefont{Goriely}},
  \bibinfo{journal}{Physics Reports} \textbf{\bibinfo{volume}{384}},
  \bibinfo{pages}{1} (\bibinfo{year}{2003}).

\bibitem[{\citenamefont{{T. Rauscher, N. Dauphas, I. Dillmann, C. Fr\"ohlich,
  Zs. F\"ul\"op and Gy. Gy\"urky}}(2013)}]{rauscher2013}
\bibinfo{author}{\bibnamefont{{T. Rauscher, N. Dauphas, I. Dillmann, C.
  Fr\"ohlich, Zs. F\"ul\"op and Gy. Gy\"urky}}}, \bibinfo{journal}{Rep. Prog.
  Phys} \textbf{\bibinfo{volume}{76}}, \bibinfo{pages}{066201}
  (\bibinfo{year}{2013}).

\bibitem[{\citenamefont{{T. Rauscher}}(2006)}]{rauscher2006}
\bibinfo{author}{\bibnamefont{{T. Rauscher}}}, \bibinfo{journal}{Phys. Rev. C}
  \textbf{\bibinfo{volume}{73}}, \bibinfo{pages}{015804}
  (\bibinfo{year}{2006}).

\bibitem[{\citenamefont{{W. Rapp, J. G\"orres, M. Wiescher, H. Schatz, and F.
  K\"appeler}}(2006)}]{rapp2006}
\bibinfo{author}{\bibnamefont{{W. Rapp, J. G\"orres, M. Wiescher, H. Schatz,
  and F. K\"appeler}}}, \bibinfo{journal}{Astroph. Journ.}
  \textbf{\bibinfo{volume}{653}}, \bibinfo{pages}{474} (\bibinfo{year}{2006}).

\bibitem[{\citenamefont{{T. Rauscher, N.Nishimura, R.Hirschi, G.Cescutti, A.
  St. J.Murphy, and A.Heger}}(2016)}]{rauscher2016}
\bibinfo{author}{\bibnamefont{{T. Rauscher, N.Nishimura, R.Hirschi, G.Cescutti,
  A. St. J.Murphy, and A.Heger}}}, \bibinfo{journal}{Mon. Not. R. Astron. Soc.}
  \textbf{\bibinfo{volume}{463}}, \bibinfo{pages}{4153} (\bibinfo{year}{2016}).

\bibitem[{\citenamefont{{N.Nishimura, T. Rauscher, R.Hirschi, G.Cescutti, A.
  St. J.Murphy, and C. Fr\"ohlich}}(2019)}]{nishimura2019}
\bibinfo{author}{\bibnamefont{{N.Nishimura, T. Rauscher, R.Hirschi, G.Cescutti,
  A. St. J.Murphy, and C. Fr\"ohlich}}}, \bibinfo{journal}{Mon. Not. R. Astron.
  Soc.} \textbf{\bibinfo{volume}{489}}, \bibinfo{pages}{1379}
  (\bibinfo{year}{2019}).

\bibitem[{\citenamefont{Hauser and Feshbach}(1952)}]{hf52}
\bibinfo{author}{\bibfnamefont{W.}~\bibnamefont{Hauser}} \bibnamefont{and}
  \bibinfo{author}{\bibfnamefont{H.}~\bibnamefont{Feshbach}},
  \bibinfo{journal}{Phys. Rev.} \textbf{\bibinfo{volume}{87}},
  \bibinfo{pages}{366} (\bibinfo{year}{1952}).

\bibitem[{\citenamefont{Koning and Delaroche}(2003)}]{Koning03}
\bibinfo{author}{\bibfnamefont{A.~J.} \bibnamefont{Koning}} \bibnamefont{and}
  \bibinfo{author}{\bibfnamefont{J.~P.} \bibnamefont{Delaroche}},
  \bibinfo{journal}{Nuclear Physics A} \textbf{\bibinfo{volume}{713}},
  \bibinfo{pages}{231} (\bibinfo{year}{2003}).

\bibitem[{\citenamefont{Bauge et~al.}(2001)\citenamefont{Bauge, Delaroche, and
  Girod}}]{bauge2001}
\bibinfo{author}{\bibfnamefont{E.}~\bibnamefont{Bauge}},
  \bibinfo{author}{\bibfnamefont{J.~P.} \bibnamefont{Delaroche}},
  \bibnamefont{and} \bibinfo{author}{\bibfnamefont{M.}~\bibnamefont{Girod}},
  \bibinfo{journal}{Phys. Rev. C} \textbf{\bibinfo{volume}{63}},
  \bibinfo{pages}{024607} (\bibinfo{year}{2001}).

\bibitem[{\citenamefont{Bauge et~al.}(1998)\citenamefont{Bauge, Delaroche, and
  Girod}}]{bauge1998}
\bibinfo{author}{\bibfnamefont{E.}~\bibnamefont{Bauge}},
  \bibinfo{author}{\bibfnamefont{J.~P.} \bibnamefont{Delaroche}},
  \bibnamefont{and} \bibinfo{author}{\bibfnamefont{M.}~\bibnamefont{Girod}},
  \bibinfo{journal}{Phys. Rev. C} \textbf{\bibinfo{volume}{58}},
  \bibinfo{pages}{1118} (\bibinfo{year}{1998}).

\bibitem[{Tal()}]{Talys1.95}
\emph{\bibinfo{title}{Nuclear reaction code \textsc{TALYS-1.95}}},
  \bibinfo{note}{(available online at: \url{http://www.talys.eu/home}}.

\bibitem[{\citenamefont{Otuka et~al.}(2014)\citenamefont{Otuka, Dupont, and
  et~al}}]{exfor}
\bibinfo{author}{\bibfnamefont{N.}~\bibnamefont{Otuka}},
  \bibinfo{author}{\bibfnamefont{E.}~\bibnamefont{Dupont}}, \bibnamefont{and}
  \bibinfo{author}{\bibfnamefont{V.~S.} \bibnamefont{et~al}},
  \bibinfo{journal}{Nuclear Data Sheets} \textbf{\bibinfo{volume}{120}},
  \bibinfo{pages}{272} (\bibinfo{year}{2014}).

\bibitem[{\citenamefont{Kennett et~al.}(1981)\citenamefont{Kennett, Mitchell,
  Anderson, and Sargood}}]{kennett81}
\bibinfo{author}{\bibfnamefont{S.~R.} \bibnamefont{Kennett}},
  \bibinfo{author}{\bibfnamefont{L.~W.} \bibnamefont{Mitchell}},
  \bibinfo{author}{\bibfnamefont{M.~R.} \bibnamefont{Anderson}},
  \bibnamefont{and} \bibinfo{author}{\bibfnamefont{D.~G.}
  \bibnamefont{Sargood}}, \bibinfo{journal}{Nuclear Physics A}
  \textbf{\bibinfo{volume}{368}}, \bibinfo{pages}{337} (\bibinfo{year}{1981}).

\bibitem[{\citenamefont{Fedorets et~al.}(1986)\citenamefont{Fedorets,
  Zalyubovsky, Nemashkalo, and Storizhko}}]{fedorets1986}
\bibinfo{author}{\bibfnamefont{I.~D.} \bibnamefont{Fedorets}},
  \bibinfo{author}{\bibfnamefont{I.~I.} \bibnamefont{Zalyubovsky}},
  \bibinfo{author}{\bibfnamefont{B.~A.} \bibnamefont{Nemashkalo}},
  \bibnamefont{and} \bibinfo{author}{\bibfnamefont{V.~E.}
  \bibnamefont{Storizhko}}, \bibinfo{journal}{Bull.Russian Academy of Sciences
  - Physics} \textbf{\bibinfo{volume}{50}}, \bibinfo{pages}{143}
  (\bibinfo{year}{1986}).

\bibitem[{\citenamefont{Kennett
  et~al.}(1980{\natexlab{a}})\citenamefont{Kennett, Anderson, Switkowski, and
  Sargood}}]{kennett80}
\bibinfo{author}{\bibfnamefont{S.~R.} \bibnamefont{Kennett}},
  \bibinfo{author}{\bibfnamefont{M.~R.} \bibnamefont{Anderson}},
  \bibinfo{author}{\bibfnamefont{Z.~E.} \bibnamefont{Switkowski}},
  \bibnamefont{and} \bibinfo{author}{\bibfnamefont{D.~G.}
  \bibnamefont{Sargood}}, \bibinfo{journal}{Nuclear Physics A}
  \textbf{\bibinfo{volume}{344}}, \bibinfo{pages}{351}
  (\bibinfo{year}{1980}{\natexlab{a}}).

\bibitem[{\citenamefont{Zyskind et~al.}(1980)\citenamefont{Zyskind, Barnes,
  Davidson, Fowler, Marrs, and Shapiro}}]{zyskind80}
\bibinfo{author}{\bibfnamefont{J.~L.} \bibnamefont{Zyskind}},
  \bibinfo{author}{\bibfnamefont{C.~A.} \bibnamefont{Barnes}},
  \bibinfo{author}{\bibfnamefont{J.~M.} \bibnamefont{Davidson}},
  \bibinfo{author}{\bibfnamefont{W.~A.} \bibnamefont{Fowler}},
  \bibinfo{author}{\bibfnamefont{R.}~\bibnamefont{Marrs}}, \bibnamefont{and}
  \bibinfo{author}{\bibfnamefont{M.}~\bibnamefont{Shapiro}},
  \bibinfo{journal}{Nuclear Physics A} \textbf{\bibinfo{volume}{343}},
  \bibinfo{pages}{295} (\bibinfo{year}{1980}).

\bibitem[{\citenamefont{Gardner et~al.}(1981)\citenamefont{Gardner, Mitchell,
  Kennett, Anderson, and Sargood}}]{gardner81}
\bibinfo{author}{\bibfnamefont{H.~J.} \bibnamefont{Gardner}},
  \bibinfo{author}{\bibfnamefont{L.~W.} \bibnamefont{Mitchell}},
  \bibinfo{author}{\bibfnamefont{S.~R.} \bibnamefont{Kennett}},
  \bibinfo{author}{\bibfnamefont{M.~R.} \bibnamefont{Anderson}},
  \bibnamefont{and} \bibinfo{author}{\bibfnamefont{D.~G.}
  \bibnamefont{Sargood}}, \bibinfo{journal}{Australian Journal of Physics}
  \textbf{\bibinfo{volume}{34}}, \bibinfo{pages}{25} (\bibinfo{year}{1981}).

\bibitem[{\citenamefont{Zyskind et~al.}(1978)\citenamefont{Zyskind, Davidson,
  Esat, Shapiro, and Spear}}]{zyskind78}
\bibinfo{author}{\bibfnamefont{J.~L.} \bibnamefont{Zyskind}},
  \bibinfo{author}{\bibfnamefont{J.~M.} \bibnamefont{Davidson}},
  \bibinfo{author}{\bibfnamefont{M.~T.} \bibnamefont{Esat}},
  \bibinfo{author}{\bibfnamefont{M.~H.} \bibnamefont{Shapiro}},
  \bibnamefont{and} \bibinfo{author}{\bibfnamefont{R.~H.} \bibnamefont{Spear}},
  \bibinfo{journal}{Nuclear Physics A} \textbf{\bibinfo{volume}{301}},
  \bibinfo{pages}{179} (\bibinfo{year}{1978}).

\bibitem[{\citenamefont{Tims et~al.}(1993)\citenamefont{Tims, Scott, Morton,
  Hansper, and Sargood}}]{tims93}
\bibinfo{author}{\bibfnamefont{S.~G.} \bibnamefont{Tims}},
  \bibinfo{author}{\bibfnamefont{A.~F.} \bibnamefont{Scott}},
  \bibinfo{author}{\bibfnamefont{A.~J.} \bibnamefont{Morton}},
  \bibinfo{author}{\bibfnamefont{V.~Y.} \bibnamefont{Hansper}},
  \bibnamefont{and} \bibinfo{author}{\bibfnamefont{D.~G.}
  \bibnamefont{Sargood}}, \bibinfo{journal}{Nuclear Physics A}
  \textbf{\bibinfo{volume}{563}}, \bibinfo{pages}{473} (\bibinfo{year}{1993}).

\bibitem[{\citenamefont{Butler and Gossett}(1957)}]{butler57}
\bibinfo{author}{\bibfnamefont{J.~W.} \bibnamefont{Butler}} \bibnamefont{and}
  \bibinfo{author}{\bibfnamefont{C.~R.} \bibnamefont{Gossett}},
  \bibinfo{journal}{Phys. Rev.} \textbf{\bibinfo{volume}{108}},
  \bibinfo{pages}{1473} (\bibinfo{year}{1957}).

\bibitem[{\citenamefont{Tingwell et~al.}(1989)\citenamefont{Tingwell, Hansper,
  Tims, Scott, Morton, and Sargood}}]{tingwell89}
\bibinfo{author}{\bibfnamefont{C.~I.~W.} \bibnamefont{Tingwell}},
  \bibinfo{author}{\bibfnamefont{V.~Y.} \bibnamefont{Hansper}},
  \bibinfo{author}{\bibfnamefont{S.~G.} \bibnamefont{Tims}},
  \bibinfo{author}{\bibfnamefont{A.~F.} \bibnamefont{Scott}},
  \bibinfo{author}{\bibfnamefont{A.~J.} \bibnamefont{Morton}},
  \bibnamefont{and} \bibinfo{author}{\bibfnamefont{D.~G.}
  \bibnamefont{Sargood}}, \bibinfo{journal}{Nuclear Physics A}
  \textbf{\bibinfo{volume}{496}}, \bibinfo{pages}{127} (\bibinfo{year}{1989}).

\bibitem[{\citenamefont{Tingwell et~al.}(1988)\citenamefont{Tingwell, Hansper,
  Tims, Scott, and Sargood}}]{tingwell88}
\bibinfo{author}{\bibfnamefont{C.~I.~W.} \bibnamefont{Tingwell}},
  \bibinfo{author}{\bibfnamefont{V.~Y.} \bibnamefont{Hansper}},
  \bibinfo{author}{\bibfnamefont{S.~G.} \bibnamefont{Tims}},
  \bibinfo{author}{\bibfnamefont{A.~F.} \bibnamefont{Scott}}, \bibnamefont{and}
  \bibinfo{author}{\bibfnamefont{D.~G.} \bibnamefont{Sargood}},
  \bibinfo{journal}{Nuclear Physics A} \textbf{\bibinfo{volume}{480}},
  \bibinfo{pages}{162} (\bibinfo{year}{1988}).

\bibitem[{\citenamefont{Sevior et~al.}(1983)\citenamefont{Sevior, Mitchell,
  Anderson, Tingwell, and Sargood}}]{sevior83}
\bibinfo{author}{\bibfnamefont{M.~E.} \bibnamefont{Sevior}},
  \bibinfo{author}{\bibfnamefont{L.~W.} \bibnamefont{Mitchell}},
  \bibinfo{author}{\bibfnamefont{M.~R.} \bibnamefont{Anderson}},
  \bibinfo{author}{\bibfnamefont{C.~I.~W.} \bibnamefont{Tingwell}},
  \bibnamefont{and} \bibinfo{author}{\bibfnamefont{D.~G.}
  \bibnamefont{Sargood}}, \bibinfo{journal}{Australian Journal of Physics}
  \textbf{\bibinfo{volume}{36}}, \bibinfo{pages}{463} (\bibinfo{year}{1983}).

\bibitem[{\citenamefont{Quinn et~al.}(2013)\citenamefont{Quinn, Spyrou, Simon,
  Battaglia, Couder, DeYoung, Dombos, Fang, Gorres, Kontos et~al.}}]{quinn2013}
\bibinfo{author}{\bibfnamefont{S.~J.} \bibnamefont{Quinn}},
  \bibinfo{author}{\bibfnamefont{A.}~\bibnamefont{Spyrou}},
  \bibinfo{author}{\bibfnamefont{A.}~\bibnamefont{Simon}},
  \bibinfo{author}{\bibfnamefont{A.}~\bibnamefont{Battaglia}},
  \bibinfo{author}{\bibfnamefont{M.}~\bibnamefont{Couder}},
  \bibinfo{author}{\bibfnamefont{P.~A.} \bibnamefont{DeYoung}},
  \bibinfo{author}{\bibfnamefont{A.~C.} \bibnamefont{Dombos}},
  \bibinfo{author}{\bibfnamefont{X.}~\bibnamefont{Fang}},
  \bibinfo{author}{\bibfnamefont{J.}~\bibnamefont{Gorres}},
  \bibinfo{author}{\bibfnamefont{A.}~\bibnamefont{Kontos}},
  \bibnamefont{et~al.}, \bibinfo{journal}{Phys. Rev. C}
  \textbf{\bibinfo{volume}{88}}, \bibinfo{pages}{011603}
  (\bibinfo{year}{2013}).

\bibitem[{\citenamefont{Krivonosov et~al.}(1977)\citenamefont{Krivonosov,
  Ekhichev, Nemashkalo, Storizhko, and Chirt}}]{krivonosov1977}
\bibinfo{author}{\bibfnamefont{G.~A.} \bibnamefont{Krivonosov}},
  \bibinfo{author}{\bibfnamefont{O.~I.} \bibnamefont{Ekhichev}},
  \bibinfo{author}{\bibfnamefont{B.~A.} \bibnamefont{Nemashkalo}},
  \bibinfo{author}{\bibfnamefont{V.~E.} \bibnamefont{Storizhko}},
  \bibnamefont{and} \bibinfo{author}{\bibfnamefont{V.~K.} \bibnamefont{Chirt}},
  \bibinfo{journal}{Bull.Russian Academy of Sciences - Physics}
  \textbf{\bibinfo{volume}{41}}, \bibinfo{pages}{333} (\bibinfo{year}{1977}).

\bibitem[{\citenamefont{Gy{\"u}rky et~al.}(2001)\citenamefont{Gy{\"u}rky,
  Somorjai, F{\"u}l{\"o}p, S.Harissopulos, P.Demetriou, and
  T.Rauscher}}]{qyurky2001}
\bibinfo{author}{\bibfnamefont{G.}~\bibnamefont{Gy{\"u}rky}},
  \bibinfo{author}{\bibfnamefont{E.}~\bibnamefont{Somorjai}},
  \bibinfo{author}{\bibfnamefont{Z.}~\bibnamefont{F{\"u}l{\"o}p}},
  \bibinfo{author}{\bibnamefont{S.Harissopulos}},
  \bibinfo{author}{\bibnamefont{P.Demetriou}}, \bibnamefont{and}
  \bibinfo{author}{\bibnamefont{T.Rauscher}}, \bibinfo{journal}{Phys. Rev. C}
  \textbf{\bibinfo{volume}{64}}, \bibinfo{pages}{065803}
  (\bibinfo{year}{2001}).

\bibitem[{\citenamefont{Galanopoulos et~al.}(2003)\citenamefont{Galanopoulos,
  Demetriou, Kokkoris, Harissopulos, Kunz, Fey, Hammer, Gy{\"u}rky,
  F{\"u}l{\"o}p, Somorjai et~al.}}]{galanopoulos2003}
\bibinfo{author}{\bibfnamefont{S.}~\bibnamefont{Galanopoulos}},
  \bibinfo{author}{\bibfnamefont{P.}~\bibnamefont{Demetriou}},
  \bibinfo{author}{\bibfnamefont{M.}~\bibnamefont{Kokkoris}},
  \bibinfo{author}{\bibfnamefont{S.}~\bibnamefont{Harissopulos}},
  \bibinfo{author}{\bibfnamefont{R.}~\bibnamefont{Kunz}},
  \bibinfo{author}{\bibfnamefont{M.}~\bibnamefont{Fey}},
  \bibinfo{author}{\bibfnamefont{J.~W.} \bibnamefont{Hammer}},
  \bibinfo{author}{\bibfnamefont{G.}~\bibnamefont{Gy{\"u}rky}},
  \bibinfo{author}{\bibfnamefont{Z.}~\bibnamefont{F{\"u}l{\"o}p}},
  \bibinfo{author}{\bibfnamefont{E.}~\bibnamefont{Somorjai}},
  \bibnamefont{et~al.}, \bibinfo{journal}{Phys. Rev. C}
  \textbf{\bibinfo{volume}{67}}, \bibinfo{pages}{015801}
  (\bibinfo{year}{2003}).

\bibitem[{\citenamefont{Harissopulos et~al.}(2013)\citenamefont{Harissopulos,
  Spyrou, Lagoyannis, Axiotis, Demetriou, Hammer, Kunz, and
  Becker}}]{harissopulos13}
\bibinfo{author}{\bibfnamefont{S.}~\bibnamefont{Harissopulos}},
  \bibinfo{author}{\bibfnamefont{A.}~\bibnamefont{Spyrou}},
  \bibinfo{author}{\bibfnamefont{A.}~\bibnamefont{Lagoyannis}},
  \bibinfo{author}{\bibfnamefont{M.}~\bibnamefont{Axiotis}},
  \bibinfo{author}{\bibfnamefont{P.}~\bibnamefont{Demetriou}},
  \bibinfo{author}{\bibfnamefont{J.~W.} \bibnamefont{Hammer}},
  \bibinfo{author}{\bibfnamefont{R.}~\bibnamefont{Kunz}}, \bibnamefont{and}
  \bibinfo{author}{\bibfnamefont{H.~W.} \bibnamefont{Becker}},
  \bibinfo{journal}{Phys. Rev. C} \textbf{\bibinfo{volume}{87}},
  \bibinfo{pages}{025806} (\bibinfo{year}{2013}).

\bibitem[{\citenamefont{Tsagari et~al.}(2004)\citenamefont{Tsagari, Kokkoris,
  Skreti, Karydas, Harissopulos, Paradellis, and Demetriou}}]{tsagari04}
\bibinfo{author}{\bibfnamefont{P.}~\bibnamefont{Tsagari}},
  \bibinfo{author}{\bibfnamefont{M.}~\bibnamefont{Kokkoris}},
  \bibinfo{author}{\bibfnamefont{E.}~\bibnamefont{Skreti}},
  \bibinfo{author}{\bibfnamefont{A.~G.} \bibnamefont{Karydas}},
  \bibinfo{author}{\bibfnamefont{S.}~\bibnamefont{Harissopulos}},
  \bibinfo{author}{\bibfnamefont{T.}~\bibnamefont{Paradellis}},
  \bibnamefont{and}
  \bibinfo{author}{\bibfnamefont{P.}~\bibnamefont{Demetriou}},
  \bibinfo{journal}{Phys. Rev. C} \textbf{\bibinfo{volume}{70}},
  \bibinfo{pages}{015802} (\bibinfo{year}{2004}).

\bibitem[{\citenamefont{Spyrou et~al.}(2013)\citenamefont{Spyrou, Quinn, Simon,
  Rauscher, Battaglia, Best, Bucher, Couder, DeYoung, Dombos
  et~al.}}]{spyrou2013}
\bibinfo{author}{\bibfnamefont{A.}~\bibnamefont{Spyrou}},
  \bibinfo{author}{\bibfnamefont{S.~J.} \bibnamefont{Quinn}},
  \bibinfo{author}{\bibfnamefont{A.}~\bibnamefont{Simon}},
  \bibinfo{author}{\bibfnamefont{T.}~\bibnamefont{Rauscher}},
  \bibinfo{author}{\bibfnamefont{A.}~\bibnamefont{Battaglia}},
  \bibinfo{author}{\bibfnamefont{A.}~\bibnamefont{Best}},
  \bibinfo{author}{\bibfnamefont{B.}~\bibnamefont{Bucher}},
  \bibinfo{author}{\bibfnamefont{M.}~\bibnamefont{Couder}},
  \bibinfo{author}{\bibfnamefont{P.~A.} \bibnamefont{DeYoung}},
  \bibinfo{author}{\bibfnamefont{A.~C.} \bibnamefont{Dombos}},
  \bibnamefont{et~al.}, \bibinfo{journal}{Phys. Rev. C}
  \textbf{\bibinfo{volume}{88}}, \bibinfo{pages}{045802}
  (\bibinfo{year}{2013}).

\bibitem[{\citenamefont{Sauter and Kappeler}(1997)}]{sauter97}
\bibinfo{author}{\bibfnamefont{T.}~\bibnamefont{Sauter}} \bibnamefont{and}
  \bibinfo{author}{\bibfnamefont{F.}~\bibnamefont{Kappeler}},
  \bibinfo{journal}{Phys. Rev. C} \textbf{\bibinfo{volume}{55}},
  \bibinfo{pages}{3127} (\bibinfo{year}{1997}).

\bibitem[{\citenamefont{Foteinou et~al.}(2019)\citenamefont{Foteinou, Axiotis,
  Harissopulos, Dimitriou, Provatas, Lagoyannis, Becker, Rogalla, Zilges,
  Schreckling et~al.}}]{foteinou2019}
\bibinfo{author}{\bibfnamefont{V.}~\bibnamefont{Foteinou}},
  \bibinfo{author}{\bibfnamefont{M.}~\bibnamefont{Axiotis}},
  \bibinfo{author}{\bibfnamefont{S.}~\bibnamefont{Harissopulos}},
  \bibinfo{author}{\bibfnamefont{P.}~\bibnamefont{Dimitriou}},
  \bibinfo{author}{\bibfnamefont{G.}~\bibnamefont{Provatas}},
  \bibinfo{author}{\bibfnamefont{A.}~\bibnamefont{Lagoyannis}},
  \bibinfo{author}{\bibfnamefont{H.}~\bibnamefont{Becker}},
  \bibinfo{author}{\bibfnamefont{D.}~\bibnamefont{Rogalla}},
  \bibinfo{author}{\bibfnamefont{A.}~\bibnamefont{Zilges}},
  \bibinfo{author}{\bibfnamefont{A.}~\bibnamefont{Schreckling}},
  \bibnamefont{et~al.}, \bibinfo{journal}{European Physical Journal A}
  \textbf{\bibinfo{volume}{55}}, \bibinfo{pages}{67} (\bibinfo{year}{2019}).

\bibitem[{\citenamefont{{J. Bork and H. Schatz and F. Kappeler and T.
  Rauscher}}(1998)}]{bork1998}
\bibinfo{author}{\bibnamefont{{J. Bork and H. Schatz and F. Kappeler and T.
  Rauscher}}}, \bibinfo{journal}{Phys. Rev. C} \textbf{\bibinfo{volume}{58}},
  \bibinfo{pages}{524} (\bibinfo{year}{1998}).

\bibitem[{\citenamefont{Dillmann et~al.}(2011)\citenamefont{Dillmann, Coquard,
  Domingo-Pardo, Kappeler, Marganiec, Uberseder, Giesen, Heiske, Feinberg,
  Hentschel et~al.}}]{dillmann2011}
\bibinfo{author}{\bibfnamefont{I.}~\bibnamefont{Dillmann}},
  \bibinfo{author}{\bibfnamefont{L.}~\bibnamefont{Coquard}},
  \bibinfo{author}{\bibfnamefont{C.}~\bibnamefont{Domingo-Pardo}},
  \bibinfo{author}{\bibfnamefont{F.}~\bibnamefont{Kappeler}},
  \bibinfo{author}{\bibfnamefont{J.}~\bibnamefont{Marganiec}},
  \bibinfo{author}{\bibfnamefont{E.}~\bibnamefont{Uberseder}},
  \bibinfo{author}{\bibfnamefont{U.}~\bibnamefont{Giesen}},
  \bibinfo{author}{\bibfnamefont{A.}~\bibnamefont{Heiske}},
  \bibinfo{author}{\bibfnamefont{G.}~\bibnamefont{Feinberg}},
  \bibinfo{author}{\bibfnamefont{D.}~\bibnamefont{Hentschel}},
  \bibnamefont{et~al.}, \bibinfo{journal}{Phys. Rev. C}
  \textbf{\bibinfo{volume}{84}}, \bibinfo{pages}{015802}
  (\bibinfo{year}{2011}).

\bibitem[{\citenamefont{Spyrou et~al.}(2008)\citenamefont{Spyrou, Lagoyannis,
  Demetriou, Harissopulos, and Becker}}]{spyrou08}
\bibinfo{author}{\bibfnamefont{A.}~\bibnamefont{Spyrou}},
  \bibinfo{author}{\bibfnamefont{A.}~\bibnamefont{Lagoyannis}},
  \bibinfo{author}{\bibfnamefont{P.}~\bibnamefont{Demetriou}},
  \bibinfo{author}{\bibfnamefont{S.}~\bibnamefont{Harissopulos}},
  \bibnamefont{and} \bibinfo{author}{\bibfnamefont{H.~W.}
  \bibnamefont{Becker}}, \bibinfo{journal}{Phys. Rev. C}
  \textbf{\bibinfo{volume}{77}}, \bibinfo{pages}{065801}
  (\bibinfo{year}{2008}).

\bibitem[{\citenamefont{Gy{\"u}rky et~al.}(2007)\citenamefont{Gy{\"u}rky, Kiss,
  Elekes, F{\"u}l{\"o}p, Somorjai, and Rauscher}}]{gyurky07}
\bibinfo{author}{\bibfnamefont{G.}~\bibnamefont{Gy{\"u}rky}},
  \bibinfo{author}{\bibfnamefont{G.~G.} \bibnamefont{Kiss}},
  \bibinfo{author}{\bibfnamefont{Z.}~\bibnamefont{Elekes}},
  \bibinfo{author}{\bibfnamefont{Z.}~\bibnamefont{F{\"u}l{\"o}p}},
  \bibinfo{author}{\bibfnamefont{E.}~\bibnamefont{Somorjai}}, \bibnamefont{and}
  \bibinfo{author}{\bibfnamefont{T.}~\bibnamefont{Rauscher}},
  \bibinfo{journal}{Journal of Physics G: Nuclear and Particle Physics}
  \textbf{\bibinfo{volume}{34}}, \bibinfo{pages}{817} (\bibinfo{year}{2007}).

\bibitem[{\citenamefont{G{\"u}ray et~al.}(2009)\citenamefont{G{\"u}ray,
  {\"O}zkan, Yal{c}in, Palumbo, deBoer, G{\"o}rres, Leblanc, Brien, Strandberg,
  Tan et~al.}}]{guray09}
\bibinfo{author}{\bibfnamefont{R.~T.} \bibnamefont{G{\"u}ray}},
  \bibinfo{author}{\bibfnamefont{N.}~\bibnamefont{{\"O}zkan}},
  \bibinfo{author}{\bibfnamefont{C.}~\bibnamefont{Yal{c}in}},
  \bibinfo{author}{\bibfnamefont{A.}~\bibnamefont{Palumbo}},
  \bibinfo{author}{\bibfnamefont{R.}~\bibnamefont{deBoer}},
  \bibinfo{author}{\bibfnamefont{J.}~\bibnamefont{G{\"o}rres}},
  \bibinfo{author}{\bibfnamefont{P.~J.} \bibnamefont{Leblanc}},
  \bibinfo{author}{\bibfnamefont{S.~O.} \bibnamefont{Brien}},
  \bibinfo{author}{\bibfnamefont{E.}~\bibnamefont{Strandberg}},
  \bibinfo{author}{\bibfnamefont{W.~P.} \bibnamefont{Tan}},
  \bibnamefont{et~al.}, \bibinfo{journal}{Phys. Rev. C}
  \textbf{\bibinfo{volume}{80}}, \bibinfo{pages}{035804}
  (\bibinfo{year}{2009}).

\bibitem[{\citenamefont{Netterdon et~al.}(2014)\citenamefont{Netterdon, Endres,
  Kiss, Mayer, Rauscher, Scholz, Sonnabend, T{\"o}r{\"o}k, and
  Zilges}}]{netterdon2014}
\bibinfo{author}{\bibfnamefont{L.}~\bibnamefont{Netterdon}},
  \bibinfo{author}{\bibfnamefont{A.}~\bibnamefont{Endres}},
  \bibinfo{author}{\bibfnamefont{G.~G.} \bibnamefont{Kiss}},
  \bibinfo{author}{\bibfnamefont{J.}~\bibnamefont{Mayer}},
  \bibinfo{author}{\bibfnamefont{T.}~\bibnamefont{Rauscher}},
  \bibinfo{author}{\bibfnamefont{P.}~\bibnamefont{Scholz}},
  \bibinfo{author}{\bibfnamefont{K.}~\bibnamefont{Sonnabend}},
  \bibinfo{author}{\bibfnamefont{Z.}~\bibnamefont{T{\"o}r{\"o}k}},
  \bibnamefont{and} \bibinfo{author}{\bibfnamefont{A.}~\bibnamefont{Zilges}},
  \bibinfo{journal}{Phys. Rev. C} \textbf{\bibinfo{volume}{90}},
  \bibinfo{pages}{035806} (\bibinfo{year}{2014}).

\bibitem[{\citenamefont{Guray et~al.}(2015)\citenamefont{Guray, {\"O}zkan,
  Yalcin, Rauscher, Gy{\"u}rky, Farkas, F{\"u}l{\"o}p, Halasz, and
  Somorjai}}]{guray2015}
\bibinfo{author}{\bibfnamefont{R.~T.} \bibnamefont{Guray}},
  \bibinfo{author}{\bibfnamefont{N.}~\bibnamefont{{\"O}zkan}},
  \bibinfo{author}{\bibfnamefont{C.}~\bibnamefont{Yalcin}},
  \bibinfo{author}{\bibfnamefont{T.}~\bibnamefont{Rauscher}},
  \bibinfo{author}{\bibfnamefont{G.}~\bibnamefont{Gy{\"u}rky}},
  \bibinfo{author}{\bibfnamefont{J.}~\bibnamefont{Farkas}},
  \bibinfo{author}{\bibfnamefont{Z.}~\bibnamefont{F{\"u}l{\"o}p}},
  \bibinfo{author}{\bibfnamefont{Z.}~\bibnamefont{Halasz}}, \bibnamefont{and}
  \bibinfo{author}{\bibfnamefont{E.}~\bibnamefont{Somorjai}},
  \bibinfo{journal}{Phys. Rev. C} \textbf{\bibinfo{volume}{91}},
  \bibinfo{pages}{055809} (\bibinfo{year}{2015}).

\bibitem[{\citenamefont{{\"O}zkan et~al.}(2017)\citenamefont{{\"O}zkan, Guray,
  Yalcin, Tan, Aprahamian, Beard, deBoer, Almaraz-Calderon, Falahat, Gorres
  et~al.}}]{ozkan2017}
\bibinfo{author}{\bibfnamefont{N.}~\bibnamefont{{\"O}zkan}},
  \bibinfo{author}{\bibfnamefont{R.~T.} \bibnamefont{Guray}},
  \bibinfo{author}{\bibfnamefont{C.}~\bibnamefont{Yalcin}},
  \bibinfo{author}{\bibfnamefont{W.~P.} \bibnamefont{Tan}},
  \bibinfo{author}{\bibfnamefont{A.}~\bibnamefont{Aprahamian}},
  \bibinfo{author}{\bibfnamefont{M.}~\bibnamefont{Beard}},
  \bibinfo{author}{\bibfnamefont{R.~J.} \bibnamefont{deBoer}},
  \bibinfo{author}{\bibfnamefont{S.}~\bibnamefont{Almaraz-Calderon}},
  \bibinfo{author}{\bibfnamefont{S.}~\bibnamefont{Falahat}},
  \bibinfo{author}{\bibfnamefont{J.}~\bibnamefont{Gorres}},
  \bibnamefont{et~al.}, \bibinfo{journal}{Phys. Rev. C}
  \textbf{\bibinfo{volume}{96}}, \bibinfo{pages}{045805}
  (\bibinfo{year}{2017}).

\bibitem[{\citenamefont{Koning et~al.}(2007)\citenamefont{Koning, Hilaire, and
  Duijvestijn}}]{cftg-talys}
\bibinfo{author}{\bibfnamefont{A.~J.} \bibnamefont{Koning}},
  \bibinfo{author}{\bibfnamefont{S.}~\bibnamefont{Hilaire}}, \bibnamefont{and}
  \bibinfo{author}{\bibfnamefont{M.~C.} \bibnamefont{Duijvestijn}}, in
  \emph{\bibinfo{booktitle}{Proceedings of the International Conference on
  Nuclear Data for Science and Technology, Nice, France}}, edited by
  \bibinfo{editor}{\bibfnamefont{O.}~\bibnamefont{Bersillon}},
  \bibinfo{editor}{\bibfnamefont{F.}~\bibnamefont{Gunsing}},
  \bibinfo{editor}{\bibfnamefont{E.}~\bibnamefont{Bauge}},
  \bibinfo{editor}{\bibfnamefont{R.}~\bibnamefont{Jacqmin}}, \bibnamefont{and}
  \bibinfo{editor}{\bibfnamefont{S.}~\bibnamefont{Leray}}
  (\bibinfo{publisher}{\textsc{EDP} Sciences}, \bibinfo{year}{2007}), p.
  \bibinfo{pages}{211}, \bibinfo{note}{available online at:
  \url{http://dx.doi.org/10.1051/ndata:07767}}.

\bibitem[{\citenamefont{Kopecky and Uhl}(1990)}]{ku1990}
\bibinfo{author}{\bibfnamefont{J.}~\bibnamefont{Kopecky}} \bibnamefont{and}
  \bibinfo{author}{\bibfnamefont{M.}~\bibnamefont{Uhl}},
  \bibinfo{journal}{Phys. Rev. C} \textbf{\bibinfo{volume}{41}},
  \bibinfo{pages}{1941} (\bibinfo{year}{1990}).

\bibitem[{\citenamefont{{P.M. Brink}}(1957)}]{brink1957}
\bibinfo{author}{\bibnamefont{{P.M. Brink}}}, \bibinfo{journal}{Nucl. Phys. A}
  \textbf{\bibinfo{volume}{4}}, \bibinfo{pages}{215} (\bibinfo{year}{1957}).

\bibitem[{\citenamefont{{P. Axel}}(1962)}]{axel1962}
\bibinfo{author}{\bibnamefont{{P. Axel}}}, \bibinfo{journal}{Phys. Rev}
  \textbf{\bibinfo{volume}{126}}, \bibinfo{pages}{671} (\bibinfo{year}{1962}).

\bibitem[{\citenamefont{{S. Goriely and E. Khan}}(2002)}]{goriely2002}
\bibinfo{author}{\bibnamefont{{S. Goriely and E. Khan}}},
  \bibinfo{journal}{Nucl. Phys. A} \textbf{\bibinfo{volume}{706}},
  \bibinfo{pages}{217} (\bibinfo{year}{2002}).

\bibitem[{\citenamefont{Goriely et~al.}(2004)\citenamefont{Goriely, Khan, and
  Samyn}}]{goriely2004}
\bibinfo{author}{\bibfnamefont{S.}~\bibnamefont{Goriely}},
  \bibinfo{author}{\bibfnamefont{E.}~\bibnamefont{Khan}}, \bibnamefont{and}
  \bibinfo{author}{\bibfnamefont{M.}~\bibnamefont{Samyn}},
  \bibinfo{journal}{Nucl. Phys. A} \textbf{\bibinfo{volume}{739}},
  \bibinfo{pages}{331} (\bibinfo{year}{2004}).

\bibitem[{\citenamefont{{S. Goriely}}(1998)}]{goriely1998}
\bibinfo{author}{\bibnamefont{{S. Goriely}}}, \bibinfo{journal}{Phys. Lett. B}
  \textbf{\bibinfo{volume}{436}}, \bibinfo{pages}{10} (\bibinfo{year}{1998}).

\bibitem[{\citenamefont{{I. Daoutidis and S. Goriely}}(2012)}]{daoutidis2012}
\bibinfo{author}{\bibnamefont{{I. Daoutidis and S. Goriely}}},
  \bibinfo{journal}{Phys. Rev. C} \textbf{\bibinfo{volume}{86}},
  \bibinfo{pages}{034328} (\bibinfo{year}{2012}).

\bibitem[{\citenamefont{{M. Martini, S. Peru, S. Hilaire, S. Goriely and F.
  Lechaftois}}(2016)}]{martini2016}
\bibinfo{author}{\bibnamefont{{M. Martini, S. Peru, S. Hilaire, S. Goriely and
  F. Lechaftois}}}, \bibinfo{journal}{Phys. Rev. C}
  \textbf{\bibinfo{volume}{94}}, \bibinfo{pages}{014304}
  (\bibinfo{year}{2016}).

\bibitem[{\citenamefont{{S. Goriely, P. Dimitriou, M. Wiedeking, \textit{et
  al.}}}(2002)}]{goriely2019}
\bibinfo{author}{\bibnamefont{{S. Goriely, P. Dimitriou, M. Wiedeking,
  \textit{et al.}}}}, \bibinfo{journal}{Eur. Phys. J. A}
  \textbf{\bibinfo{volume}{55}}, \bibinfo{pages}{172} (\bibinfo{year}{2002}).

\bibitem[{\citenamefont{{A.J. Koning, S. Hilaire and S.
  Goriely}}(2008)}]{koning2008}
\bibinfo{author}{\bibnamefont{{A.J. Koning, S. Hilaire and S. Goriely}}},
  \bibinfo{journal}{Nucl. Phys. A} \textbf{\bibinfo{volume}{810}},
  \bibinfo{pages}{13} (\bibinfo{year}{2008}).

\bibitem[{\citenamefont{{W. Dilg, W. Schantl, H. Vonach and M.
  Uhl}}(1973)}]{dilg1973}
\bibinfo{author}{\bibnamefont{{W. Dilg, W. Schantl, H. Vonach and M. Uhl}}},
  \bibinfo{journal}{Nucl. Phys. A} \textbf{\bibinfo{volume}{217}},
  \bibinfo{pages}{269} (\bibinfo{year}{1973}).

\bibitem[{\citenamefont{{A.V. Ignatyuk, K.K. Istekov, and G.N.
  Smirenkin}}(1979)}]{ignatyuk1979}
\bibinfo{author}{\bibnamefont{{A.V. Ignatyuk, K.K. Istekov, and G.N.
  Smirenkin}}}, \bibinfo{journal}{Sov. J. Nucl. Phys}
  \textbf{\bibinfo{volume}{29}}, \bibinfo{pages}{450} (\bibinfo{year}{1979}).

\bibitem[{\citenamefont{{A.V. Ignatyuk, J.L. Weil, S. Raman, and S.
  Kahane}}(1993)}]{ignatyuk1993}
\bibinfo{author}{\bibnamefont{{A.V. Ignatyuk, J.L. Weil, S. Raman, and S.
  Kahane}}}, \bibinfo{journal}{Phys. Rev. C} \textbf{\bibinfo{volume}{47}},
  \bibinfo{pages}{1504} (\bibinfo{year}{1993}).

\bibitem[{\citenamefont{{P. Demetriou and S. Goriely}}(2001)}]{demetriou2001}
\bibinfo{author}{\bibnamefont{{P. Demetriou and S. Goriely}}},
  \bibinfo{journal}{Nucl. Phys. A} \textbf{\bibinfo{volume}{695}},
  \bibinfo{pages}{95} (\bibinfo{year}{2001}).

\bibitem[{\citenamefont{{S. Goriely, S. Hilaire and A.J.
  Koning}}(2008)}]{goriely2008}
\bibinfo{author}{\bibnamefont{{S. Goriely, S. Hilaire and A.J. Koning}}},
  \bibinfo{journal}{Phys. Rev. C} \textbf{\bibinfo{volume}{78}},
  \bibinfo{pages}{064307} (\bibinfo{year}{2008}).

\bibitem[{\citenamefont{{S. Hilaire, S. Goriely and A.J.
  Koning}}(2012)}]{hilaire2013}
\bibinfo{author}{\bibnamefont{{S. Hilaire, S. Goriely and A.J. Koning}}},
  \bibinfo{journal}{Phys. Rev. C} \textbf{\bibinfo{volume}{86}},
  \bibinfo{pages}{064317} (\bibinfo{year}{2012}).

\bibitem[{\citenamefont{Blaser et~al.}(1951)\citenamefont{Blaser, Boehm,
  Marmier, and Scherrer}}]{blaser1951}
\bibinfo{author}{\bibfnamefont{J.~P.} \bibnamefont{Blaser}},
  \bibinfo{author}{\bibfnamefont{F.}~\bibnamefont{Boehm}},
  \bibinfo{author}{\bibfnamefont{P.}~\bibnamefont{Marmier}}, \bibnamefont{and}
  \bibinfo{author}{\bibfnamefont{P.}~\bibnamefont{Scherrer}},
  \bibinfo{journal}{Helvetica Physica Acta} \textbf{\bibinfo{volume}{24}},
  \bibinfo{pages}{3} (\bibinfo{year}{1951}).

\bibitem[{\citenamefont{Johnson et~al.}(1964)\citenamefont{Johnson, Trail, and
  Galonsky}}]{johnson1964}
\bibinfo{author}{\bibfnamefont{C.~H.} \bibnamefont{Johnson}},
  \bibinfo{author}{\bibfnamefont{C.~C.} \bibnamefont{Trail}}, \bibnamefont{and}
  \bibinfo{author}{\bibfnamefont{A.}~\bibnamefont{Galonsky}},
  \bibinfo{journal}{Phys. Rev.} \textbf{\bibinfo{volume}{136}},
  \bibinfo{pages}{B1719} (\bibinfo{year}{1964}).

\bibitem[{\citenamefont{Skakun et~al.}(1987)\citenamefont{Skakun, Batij,
  Rakivnenko, and Rastrepin}}]{skakun1987}
\bibinfo{author}{\bibfnamefont{E.~A.} \bibnamefont{Skakun}},
  \bibinfo{author}{\bibfnamefont{V.~G.} \bibnamefont{Batij}},
  \bibinfo{author}{\bibfnamefont{Y.~N.} \bibnamefont{Rakivnenko}},
  \bibnamefont{and} \bibinfo{author}{\bibfnamefont{O.~A.}
  \bibnamefont{Rastrepin}}, \bibinfo{journal}{Yadernaya Fizika}
  \textbf{\bibinfo{volume}{46}}, \bibinfo{pages}{28} (\bibinfo{year}{1987}).

\bibitem[{\citenamefont{Johnson et~al.}(1958)\citenamefont{Johnson, Galonsky,
  and Ulrich}}]{johnson1958}
\bibinfo{author}{\bibfnamefont{C.~H.} \bibnamefont{Johnson}},
  \bibinfo{author}{\bibfnamefont{A.}~\bibnamefont{Galonsky}}, \bibnamefont{and}
  \bibinfo{author}{\bibfnamefont{J.~P.} \bibnamefont{Ulrich}},
  \bibinfo{journal}{Phys. Rev.} \textbf{\bibinfo{volume}{109}},
  \bibinfo{pages}{1243} (\bibinfo{year}{1958}).

\bibitem[{\citenamefont{Kennett
  et~al.}(1980{\natexlab{b}})\citenamefont{Kennett, Anderson, Switkowski, and
  Sargood}}]{kennett1980}
\bibinfo{author}{\bibfnamefont{S.~R.} \bibnamefont{Kennett}},
  \bibinfo{author}{\bibfnamefont{M.~R.} \bibnamefont{Anderson}},
  \bibinfo{author}{\bibfnamefont{Z.~E.} \bibnamefont{Switkowski}},
  \bibnamefont{and} \bibinfo{author}{\bibfnamefont{D.~G.}
  \bibnamefont{Sargood}}, \bibinfo{journal}{Phys. Rev. C}
  \textbf{\bibinfo{volume}{344}}, \bibinfo{pages}{351}
  (\bibinfo{year}{1980}{\natexlab{b}}).

\bibitem[{\citenamefont{Harris et~al.}(1965)\citenamefont{Harris, Grench,
  Johnson, and Vaughn}}]{harris1965}
\bibinfo{author}{\bibfnamefont{K.~K.} \bibnamefont{Harris}},
  \bibinfo{author}{\bibfnamefont{H.~A.} \bibnamefont{Grench}},
  \bibinfo{author}{\bibfnamefont{R.~G.} \bibnamefont{Johnson}},
  \bibnamefont{and} \bibinfo{author}{\bibfnamefont{F.~J.}
  \bibnamefont{Vaughn}}, \bibinfo{journal}{Nuclear Instruments and Methods}
  \textbf{\bibinfo{volume}{33}}, \bibinfo{pages}{257} (\bibinfo{year}{1965}).

\bibitem[{\citenamefont{Mehta et~al.}(1977)\citenamefont{Mehta, Kailas, and
  Sekhara}}]{mehta1977}
\bibinfo{author}{\bibfnamefont{M.~K.} \bibnamefont{Mehta}},
  \bibinfo{author}{\bibfnamefont{S.}~\bibnamefont{Kailas}}, \bibnamefont{and}
  \bibinfo{author}{\bibfnamefont{K.~K.} \bibnamefont{Sekhara}},
  \bibinfo{journal}{Nuclear And Particle Physics} \textbf{\bibinfo{volume}{9}},
  \bibinfo{pages}{419} (\bibinfo{year}{1977}).

\bibitem[{\citenamefont{Barrandon et~al.}(1975)\citenamefont{Barrandon, Debrun,
  Kohn, and Spearh}}]{barrandon1975}
\bibinfo{author}{\bibfnamefont{J.~N.} \bibnamefont{Barrandon}},
  \bibinfo{author}{\bibfnamefont{J.~L.} \bibnamefont{Debrun}},
  \bibinfo{author}{\bibfnamefont{A.}~\bibnamefont{Kohn}}, \bibnamefont{and}
  \bibinfo{author}{\bibfnamefont{R.~H.} \bibnamefont{Spearh}},
  \bibinfo{journal}{Nuclear Instruments and Methods}
  \textbf{\bibinfo{volume}{127}}, \bibinfo{pages}{269} (\bibinfo{year}{1975}).

\bibitem[{\citenamefont{Johnson et~al.}(1960)\citenamefont{Johnson, Galonsky,
  and Inskeep}}]{johnson1960}
\bibinfo{author}{\bibfnamefont{C.~H.} \bibnamefont{Johnson}},
  \bibinfo{author}{\bibfnamefont{A.}~\bibnamefont{Galonsky}}, \bibnamefont{and}
  \bibinfo{author}{\bibfnamefont{C.~N.} \bibnamefont{Inskeep}}, vol.
  \bibinfo{volume}{109} (\bibinfo{publisher}{Prog: Oak Ridge National Lab.
  Reports}, \bibinfo{year}{1960}).

\bibitem[{\citenamefont{Kailas et~al.}(1975)\citenamefont{Kailas, Gupta, Mehta,
  Kerekatte, Namjoshi, Ganguly, and Chintalapudi}}]{kailas1975}
\bibinfo{author}{\bibfnamefont{S.}~\bibnamefont{Kailas}},
  \bibinfo{author}{\bibfnamefont{S.~K.} \bibnamefont{Gupta}},
  \bibinfo{author}{\bibfnamefont{M.~K.} \bibnamefont{Mehta}},
  \bibinfo{author}{\bibfnamefont{S.~S.} \bibnamefont{Kerekatte}},
  \bibinfo{author}{\bibfnamefont{L.~V.} \bibnamefont{Namjoshi}},
  \bibinfo{author}{\bibfnamefont{N.~K.} \bibnamefont{Ganguly}},
  \bibnamefont{and}
  \bibinfo{author}{\bibfnamefont{S.}~\bibnamefont{Chintalapudi}},
  \bibinfo{journal}{Phys. Rev. C} \textbf{\bibinfo{volume}{12}},
  \bibinfo{pages}{1789} (\bibinfo{year}{1975}).

\bibitem[{\citenamefont{Mitchell and Sargood}(1983)}]{mitchell1983}
\bibinfo{author}{\bibfnamefont{L.~W.} \bibnamefont{Mitchell}} \bibnamefont{and}
  \bibinfo{author}{\bibfnamefont{D.~G.} \bibnamefont{Sargood}},
  \bibinfo{journal}{Aust. J. Phys.} \textbf{\bibinfo{volume}{36}},
  \bibinfo{pages}{1} (\bibinfo{year}{1983}).

\bibitem[{\citenamefont{Switkowski et~al.}(1978)\citenamefont{Switkowski,
  Heggie, and Mann}}]{switkowski1978}
\bibinfo{author}{\bibfnamefont{Z.~E.} \bibnamefont{Switkowski}},
  \bibinfo{author}{\bibfnamefont{J.~C.~P.} \bibnamefont{Heggie}},
  \bibnamefont{and} \bibinfo{author}{\bibfnamefont{F.~M.} \bibnamefont{Mann}},
  \bibinfo{journal}{Aust. J. Phys.} \textbf{\bibinfo{volume}{31}},
  \bibinfo{pages}{253} (\bibinfo{year}{1978}).

\bibitem[{\citenamefont{Saini et~al.}(1983)\citenamefont{Saini, Singh,
  Chatterjee, Kailas, Karnik, Veerabahu, and Mehta}}]{saini1983}
\bibinfo{author}{\bibfnamefont{S.}~\bibnamefont{Saini}},
  \bibinfo{author}{\bibfnamefont{G.}~\bibnamefont{Singh}},
  \bibinfo{author}{\bibfnamefont{A.}~\bibnamefont{Chatterjee}},
  \bibinfo{author}{\bibfnamefont{S.}~\bibnamefont{Kailas}},
  \bibinfo{author}{\bibfnamefont{D.}~\bibnamefont{Karnik}},
  \bibinfo{author}{\bibfnamefont{N.}~\bibnamefont{Veerabahu}},
  \bibnamefont{and} \bibinfo{author}{\bibfnamefont{M.~K.} \bibnamefont{Mehta}},
  \bibinfo{journal}{Nuclear Physics A} \textbf{\bibinfo{volume}{405}},
  \bibinfo{pages}{55} (\bibinfo{year}{1983}).

\bibitem[{\citenamefont{Hershberger and Gabbard}(1984)}]{hershberger1984}
\bibinfo{author}{\bibfnamefont{R.~L.} \bibnamefont{Hershberger}}
  \bibnamefont{and} \bibinfo{author}{\bibfnamefont{F.}~\bibnamefont{Gabbard}},
  \bibinfo{journal}{Bulletin of the American Physical Society Ser.II}
  \textbf{\bibinfo{volume}{28}}, \bibinfo{pages}{734} (\bibinfo{year}{1984}).

\bibitem[{\citenamefont{Generalov et~al.}(2017)\citenamefont{Generalov,
  Abramovich, and Selyankina}}]{generalov2017}
\bibinfo{author}{\bibfnamefont{L.~N.} \bibnamefont{Generalov}},
  \bibinfo{author}{\bibfnamefont{S.~N.} \bibnamefont{Abramovich}},
  \bibnamefont{and} \bibinfo{author}{\bibfnamefont{S.~M.}
  \bibnamefont{Selyankina}}, \bibinfo{journal}{Bulletin of the Russian Academy
  of Sciences: Physics} \textbf{\bibinfo{volume}{81}}, \bibinfo{pages}{644}
  (\bibinfo{year}{2017}).

\bibitem[{\citenamefont{Kiss et~al.}(2007)\citenamefont{Kiss, Gy{\"u}rky,
  Elekes, F{\"u}l{\"o}p, Samorjai, Rauscher, and Wiescher}}]{kiss2007}
\bibinfo{author}{\bibfnamefont{G.~G.} \bibnamefont{Kiss}},
  \bibinfo{author}{\bibfnamefont{G.}~\bibnamefont{Gy{\"u}rky}},
  \bibinfo{author}{\bibfnamefont{Z.}~\bibnamefont{Elekes}},
  \bibinfo{author}{\bibfnamefont{Z.}~\bibnamefont{F{\"u}l{\"o}p}},
  \bibinfo{author}{\bibfnamefont{E.}~\bibnamefont{Samorjai}},
  \bibinfo{author}{\bibfnamefont{T.}~\bibnamefont{Rauscher}}, \bibnamefont{and}
  \bibinfo{author}{\bibfnamefont{M.}~\bibnamefont{Wiescher}},
  \bibinfo{journal}{Phys. Rev. C} \textbf{\bibinfo{volume}{76}},
  \bibinfo{pages}{055807} (\bibinfo{year}{2007}).

\bibitem[{\citenamefont{Albert}(1959)}]{albert1959}
\bibinfo{author}{\bibfnamefont{R.}~\bibnamefont{Albert}},
  \bibinfo{journal}{Phys. Rev.} \textbf{\bibinfo{volume}{115}},
  \bibinfo{pages}{925} (\bibinfo{year}{1959}).

\bibitem[{\citenamefont{Foteinou et~al.}(2018)\citenamefont{Foteinou,
  Harissopulos, Axiotis, Lagoyannis, Provatas, Spyrou, Perdikakis, Zarkadas,
  and Demetriou}}]{foteinou2018}
\bibinfo{author}{\bibfnamefont{V.}~\bibnamefont{Foteinou}},
  \bibinfo{author}{\bibfnamefont{S.}~\bibnamefont{Harissopulos}},
  \bibinfo{author}{\bibfnamefont{M.}~\bibnamefont{Axiotis}},
  \bibinfo{author}{\bibfnamefont{A.}~\bibnamefont{Lagoyannis}},
  \bibinfo{author}{\bibfnamefont{G.}~\bibnamefont{Provatas}},
  \bibinfo{author}{\bibfnamefont{A.}~\bibnamefont{Spyrou}},
  \bibinfo{author}{\bibfnamefont{G.}~\bibnamefont{Perdikakis}},
  \bibinfo{author}{\bibfnamefont{C.}~\bibnamefont{Zarkadas}}, \bibnamefont{and}
  \bibinfo{author}{\bibfnamefont{P.}~\bibnamefont{Demetriou}},
  \bibinfo{journal}{Phys. Rev. C} \textbf{\bibinfo{volume}{97}},
  \bibinfo{pages}{035806} (\bibinfo{year}{2018}).

\bibitem[{\citenamefont{Kailas et~al.}(1979)\citenamefont{Kailas, Saini, Mehta,
  Veerabahu, Viyogi, and Ganguly}}]{kailas1979}
\bibinfo{author}{\bibfnamefont{S.}~\bibnamefont{Kailas}},
  \bibinfo{author}{\bibfnamefont{S.}~\bibnamefont{Saini}},
  \bibinfo{author}{\bibfnamefont{M.~K.} \bibnamefont{Mehta}},
  \bibinfo{author}{\bibfnamefont{N.}~\bibnamefont{Veerabahu}},
  \bibinfo{author}{\bibfnamefont{Y.~P.} \bibnamefont{Viyogi}},
  \bibnamefont{and} \bibinfo{author}{\bibfnamefont{N.~K.}
  \bibnamefont{Ganguly}}, \bibinfo{journal}{Nuclear Physics A}
  \textbf{\bibinfo{volume}{315}}, \bibinfo{pages}{157} (\bibinfo{year}{1979}).

\bibitem[{\citenamefont{Debuyst and Stricht}(1968)}]{debuyst1968}
\bibinfo{author}{\bibfnamefont{R.}~\bibnamefont{Debuyst}} \bibnamefont{and}
  \bibinfo{author}{\bibfnamefont{A.~V.} \bibnamefont{Stricht}},
  \bibinfo{journal}{J.Inorg.Nucl Chem.} \textbf{\bibinfo{volume}{115}},
  \bibinfo{pages}{691} (\bibinfo{year}{1968}).

\bibitem[{\citenamefont{Gy{\"u}rky et~al.}(2003)\citenamefont{Gy{\"u}rky,
  F{\"u}l{\"o}p, Somorjai, Kokkoris, Galanopoulos, Demetriou, Harissopulos,
  Rauscher, and Goriely}}]{gyurky2003}
\bibinfo{author}{\bibfnamefont{G.}~\bibnamefont{Gy{\"u}rky}},
  \bibinfo{author}{\bibfnamefont{Z.}~\bibnamefont{F{\"u}l{\"o}p}},
  \bibinfo{author}{\bibfnamefont{E.}~\bibnamefont{Somorjai}},
  \bibinfo{author}{\bibfnamefont{M.}~\bibnamefont{Kokkoris}},
  \bibinfo{author}{\bibfnamefont{S.}~\bibnamefont{Galanopoulos}},
  \bibinfo{author}{\bibfnamefont{P.}~\bibnamefont{Demetriou}},
  \bibinfo{author}{\bibfnamefont{S.}~\bibnamefont{Harissopulos}},
  \bibinfo{author}{\bibfnamefont{T.}~\bibnamefont{Rauscher}}, \bibnamefont{and}
  \bibinfo{author}{\bibfnamefont{S.}~\bibnamefont{Goriely}},
  \bibinfo{journal}{Phys. Rev. C} \textbf{\bibinfo{volume}{68}},
  \bibinfo{pages}{055803} (\bibinfo{year}{2003}).

\bibitem[{\citenamefont{West et~al.}(1993)\citenamefont{West, Nuckolls, Hudson,
  Ruiz, Lanier, and Mustafa}}]{west1993}
\bibinfo{author}{\bibfnamefont{H.~I.} \bibnamefont{West}},
  \bibinfo{author}{\bibfnamefont{R.~M.} \bibnamefont{Nuckolls}},
  \bibinfo{author}{\bibfnamefont{B.}~\bibnamefont{Hudson}},
  \bibinfo{author}{\bibfnamefont{B.}~\bibnamefont{Ruiz}},
  \bibinfo{author}{\bibfnamefont{R.}~\bibnamefont{Lanier}}, \bibnamefont{and}
  \bibinfo{author}{\bibfnamefont{M.~G.} \bibnamefont{Mustafa}},
  \bibinfo{journal}{Phys. Rev. C} \textbf{\bibinfo{volume}{47}},
  \bibinfo{pages}{248} (\bibinfo{year}{1993}).

\bibitem[{\citenamefont{Kiss et~al.}(2008)\citenamefont{Kiss, Rauscher,
  Gy{\"u}rky, Simon, F{\"u}l{\"o}p, and Samorjai}}]{kiss2008}
\bibinfo{author}{\bibfnamefont{G.~G.} \bibnamefont{Kiss}},
  \bibinfo{author}{\bibfnamefont{T.}~\bibnamefont{Rauscher}},
  \bibinfo{author}{\bibfnamefont{G.}~\bibnamefont{Gy{\"u}rky}},
  \bibinfo{author}{\bibfnamefont{A.}~\bibnamefont{Simon}},
  \bibinfo{author}{\bibfnamefont{Z.}~\bibnamefont{F{\"u}l{\"o}p}},
  \bibnamefont{and} \bibinfo{author}{\bibfnamefont{E.}~\bibnamefont{Samorjai}},
  \bibinfo{journal}{Phys. Rev. Letters} \textbf{\bibinfo{volume}{101}},
  \bibinfo{pages}{191101} (\bibinfo{year}{2008}).

\bibitem[{\citenamefont{Flynn et~al.}(1979)\citenamefont{Flynn, Hershberger,
  and Gabbard}}]{flynn1979}
\bibinfo{author}{\bibfnamefont{D.~S.} \bibnamefont{Flynn}},
  \bibinfo{author}{\bibfnamefont{R.~L.} \bibnamefont{Hershberger}},
  \bibnamefont{and} \bibinfo{author}{\bibfnamefont{F.}~\bibnamefont{Gabbard}},
  \bibinfo{journal}{Phys. Rev. C.} \textbf{\bibinfo{volume}{20}},
  \bibinfo{pages}{1700} (\bibinfo{year}{1979}).

\bibitem[{\citenamefont{Fedorets et~al.}(1977)\citenamefont{Fedorets,
  Mishchenko, Popov, and Storizhko}}]{fedorets1977}
\bibinfo{author}{\bibfnamefont{I.~D.} \bibnamefont{Fedorets}},
  \bibinfo{author}{\bibfnamefont{V.~M.} \bibnamefont{Mishchenko}},
  \bibinfo{author}{\bibfnamefont{A.~I.} \bibnamefont{Popov}}, \bibnamefont{and}
  \bibinfo{author}{\bibfnamefont{V.~E.} \bibnamefont{Storizhko}},
  \bibinfo{journal}{Bull.Russian Academy of Sciences - Physics}
  \textbf{\bibinfo{volume}{41}}, \bibinfo{pages}{110} (\bibinfo{year}{1977}).

\bibitem[{\citenamefont{Sud{\'a}r et~al.}(2002)\citenamefont{Sud{\'a}r,
  Cserp{\'a}k, and Qaim}}]{sudar2002}
\bibinfo{author}{\bibfnamefont{S.}~\bibnamefont{Sud{\'a}r}},
  \bibinfo{author}{\bibfnamefont{F.}~\bibnamefont{Cserp{\'a}k}},
  \bibnamefont{and} \bibinfo{author}{\bibfnamefont{S.}~\bibnamefont{Qaim}},
  \bibinfo{journal}{Applied Radiation and Isotopes}
  \textbf{\bibinfo{volume}{56}}, \bibinfo{pages}{821} (\bibinfo{year}{2002}).

\bibitem[{\citenamefont{Wing and Huizenga}(1962)}]{wing1962}
\bibinfo{author}{\bibfnamefont{J.}~\bibnamefont{Wing}} \bibnamefont{and}
  \bibinfo{author}{\bibfnamefont{J.~R.} \bibnamefont{Huizenga}},
  \bibinfo{journal}{Phys. Rev.} \textbf{\bibinfo{volume}{128}},
  \bibinfo{pages}{280} (\bibinfo{year}{1962}).

\bibitem[{\citenamefont{Hershberger et~al.}(1980)\citenamefont{Hershberger,
  Flynn, Gabbard, and Johnson}}]{hershberger1980}
\bibinfo{author}{\bibfnamefont{R.~L.} \bibnamefont{Hershberger}},
  \bibinfo{author}{\bibfnamefont{D.~S.} \bibnamefont{Flynn}},
  \bibinfo{author}{\bibfnamefont{F.}~\bibnamefont{Gabbard}}, \bibnamefont{and}
  \bibinfo{author}{\bibfnamefont{C.~H.} \bibnamefont{Johnson}},
  \bibinfo{journal}{Phys. Rev. C} \textbf{\bibinfo{volume}{29}},
  \bibinfo{pages}{896} (\bibinfo{year}{1980}).

\bibitem[{\citenamefont{Dmitriev et~al.}(1967)\citenamefont{Dmitriev,
  Konstantinov, and Krasnov}}]{dmitriev1967}
\bibinfo{author}{\bibfnamefont{P.~P.} \bibnamefont{Dmitriev}},
  \bibinfo{author}{\bibfnamefont{I.~O.} \bibnamefont{Konstantinov}},
  \bibnamefont{and} \bibinfo{author}{\bibfnamefont{N.~N.}
  \bibnamefont{Krasnov}}, \bibinfo{journal}{Soviet Atomic Energy}
  \textbf{\bibinfo{volume}{22}}, \bibinfo{pages}{386} (\bibinfo{year}{1967}).

\bibitem[{\citenamefont{Skakun et~al.}(1975)\citenamefont{Skakun, Klyucharev,
  Rakivnenko, and Romanii}}]{skakun1975}
\bibinfo{author}{\bibfnamefont{E.~A.} \bibnamefont{Skakun}},
  \bibinfo{author}{\bibfnamefont{A.~P.} \bibnamefont{Klyucharev}},
  \bibinfo{author}{\bibfnamefont{Y.~N.} \bibnamefont{Rakivnenko}},
  \bibnamefont{and} \bibinfo{author}{\bibfnamefont{I.~A.}
  \bibnamefont{Romanii}}, \bibinfo{journal}{Bull.Russian Academy of
  Sciences-Physics} \textbf{\bibinfo{volume}{39}}, \bibinfo{pages}{18}
  (\bibinfo{year}{1975}).

\bibitem[{\citenamefont{Klyucharev et~al.}(1970)\citenamefont{Klyucharev,
  Skakun, Rakivnenko, and Romanii}}]{klucharev1970}
\bibinfo{author}{\bibfnamefont{A.~P.} \bibnamefont{Klyucharev}},
  \bibinfo{author}{\bibfnamefont{E.~A.} \bibnamefont{Skakun}},
  \bibinfo{author}{\bibfnamefont{Y.~N.} \bibnamefont{Rakivnenko}},
  \bibnamefont{and} \bibinfo{author}{\bibfnamefont{I.~A.}
  \bibnamefont{Romanii}}, \bibinfo{journal}{Yadernaya Fizika}
  \textbf{\bibinfo{volume}{11}}, \bibinfo{pages}{953} (\bibinfo{year}{1970}).

\bibitem[{\citenamefont{Johnson and Kernell}(1970)}]{johnson1970}
\bibinfo{author}{\bibfnamefont{C.~H.} \bibnamefont{Johnson}} \bibnamefont{and}
  \bibinfo{author}{\bibfnamefont{R.~L.} \bibnamefont{Kernell}},
  \bibinfo{journal}{Phys. Rev. C} \textbf{\bibinfo{volume}{2}},
  \bibinfo{pages}{639} (\bibinfo{year}{1970}).

\bibitem[{\citenamefont{Johnson et~al.}(1977)\citenamefont{Johnson, Bair,
  Jones, Penny, and Smith}}]{johnson1977}
\bibinfo{author}{\bibfnamefont{C.~H.} \bibnamefont{Johnson}},
  \bibinfo{author}{\bibfnamefont{J.~K.} \bibnamefont{Bair}},
  \bibinfo{author}{\bibfnamefont{C.~M.} \bibnamefont{Jones}},
  \bibinfo{author}{\bibfnamefont{S.~K.} \bibnamefont{Penny}}, \bibnamefont{and}
  \bibinfo{author}{\bibfnamefont{D.~W.} \bibnamefont{Smith}},
  \bibinfo{journal}{Phys. Rev. C} \textbf{\bibinfo{volume}{15}},
  \bibinfo{pages}{196} (\bibinfo{year}{1977}).

\bibitem[{\citenamefont{Batij and Skakun}(1991)}]{batij1991}
\bibinfo{author}{\bibfnamefont{V.~G.} \bibnamefont{Batij}} \bibnamefont{and}
  \bibinfo{author}{\bibfnamefont{E.~A.} \bibnamefont{Skakun}},
  \bibinfo{journal}{Conf.on Nucl.Data for Sci. and Technology}
  \textbf{\bibinfo{volume}{2}}, \bibinfo{pages}{1325} (\bibinfo{year}{1991}).

\bibitem[{\citenamefont{Elmaghraby et~al.}(2009)\citenamefont{Elmaghraby, Said,
  and Asfour}}]{elmaghraby2009}
\bibinfo{author}{\bibfnamefont{E.~K.} \bibnamefont{Elmaghraby}},
  \bibinfo{author}{\bibfnamefont{S.~A.} \bibnamefont{Said}}, \bibnamefont{and}
  \bibinfo{author}{\bibfnamefont{F.~I.} \bibnamefont{Asfour}},
  \bibinfo{journal}{Applied Radiation and Isotopes}
  \textbf{\bibinfo{volume}{67}}, \bibinfo{pages}{147} (\bibinfo{year}{2009}).

\bibitem[{\citenamefont{Colle and Kishore}(1974)}]{colle1974}
\bibinfo{author}{\bibfnamefont{R.}~\bibnamefont{Colle}} \bibnamefont{and}
  \bibinfo{author}{\bibfnamefont{R.}~\bibnamefont{Kishore}},
  \bibinfo{journal}{Phys. Rev. C} \textbf{\bibinfo{volume}{9}},
  \bibinfo{pages}{2166} (\bibinfo{year}{1974}).

\bibitem[{\citenamefont{Verdieck and Miller}(1967)}]{verdieck1967}
\bibinfo{author}{\bibfnamefont{E.~V.} \bibnamefont{Verdieck}} \bibnamefont{and}
  \bibinfo{author}{\bibfnamefont{J.~M.} \bibnamefont{Miller}},
  \bibinfo{journal}{Phys. Rev.} \textbf{\bibinfo{volume}{153}},
  \bibinfo{pages}{1253} (\bibinfo{year}{1967}).

\bibitem[{\citenamefont{Gheorghe et~al.}(2014)\citenamefont{Gheorghe,
  Filipescu, Glodariu, Bucurescu, Cata-Danil, Cata-Danil, Deleanu, Ghita,
  Ivascu, Lica et~al.}}]{gheorghe2014}
\bibinfo{author}{\bibfnamefont{I.}~\bibnamefont{Gheorghe}},
  \bibinfo{author}{\bibfnamefont{D.}~\bibnamefont{Filipescu}},
  \bibinfo{author}{\bibfnamefont{T.}~\bibnamefont{Glodariu}},
  \bibinfo{author}{\bibfnamefont{D.}~\bibnamefont{Bucurescu}},
  \bibinfo{author}{\bibfnamefont{I.}~\bibnamefont{Cata-Danil}},
  \bibinfo{author}{\bibfnamefont{G.}~\bibnamefont{Cata-Danil}},
  \bibinfo{author}{\bibfnamefont{D.}~\bibnamefont{Deleanu}},
  \bibinfo{author}{\bibfnamefont{D.}~\bibnamefont{Ghita}},
  \bibinfo{author}{\bibfnamefont{M.}~\bibnamefont{Ivascu}},
  \bibinfo{author}{\bibfnamefont{R.}~\bibnamefont{Lica}}, \bibnamefont{et~al.},
  \bibinfo{journal}{Nuclear Data Sheets} \textbf{\bibinfo{volume}{119}},
  \bibinfo{pages}{245} (\bibinfo{year}{2014}).

\bibitem[{\citenamefont{West et~al.}(1989)\citenamefont{West, Lanier, Mustafa,
  Nuckolls, Frehaut, Adam, and Philis}}]{west1989}
\bibinfo{author}{\bibfnamefont{H.~I.} \bibnamefont{West}},
  \bibinfo{author}{\bibfnamefont{R.~G.} \bibnamefont{Lanier}},
  \bibinfo{author}{\bibfnamefont{M.~G.} \bibnamefont{Mustafa}},
  \bibinfo{author}{\bibfnamefont{R.~N.} \bibnamefont{Nuckolls}},
  \bibinfo{author}{\bibfnamefont{J.}~\bibnamefont{Frehaut}},
  \bibinfo{author}{\bibfnamefont{A.}~\bibnamefont{Adam}}, \bibnamefont{and}
  \bibinfo{author}{\bibfnamefont{C.~A.} \bibnamefont{Philis}},
  \bibinfo{journal}{Brookhaven National Laboratory Reports}
  \textbf{\bibinfo{volume}{42382}}, \bibinfo{pages}{87} (\bibinfo{year}{1989}).

\bibitem[{\citenamefont{Spahn et~al.}(2005)\citenamefont{Spahn, Tak{\'a}cs,
  Shubin, T{\'a}rk{\'a}nyi, Coenen, and Qaim}}]{spahn2005}
\bibinfo{author}{\bibfnamefont{I.}~\bibnamefont{Spahn}},
  \bibinfo{author}{\bibfnamefont{S.}~\bibnamefont{Tak{\'a}cs}},
  \bibinfo{author}{\bibfnamefont{Y.}~\bibnamefont{Shubin}},
  \bibinfo{author}{\bibfnamefont{F.}~\bibnamefont{T{\'a}rk{\'a}nyi}},
  \bibinfo{author}{\bibfnamefont{H.}~\bibnamefont{Coenen}}, \bibnamefont{and}
  \bibinfo{author}{\bibfnamefont{S.}~\bibnamefont{Qaim}},
  \bibinfo{journal}{Applied Radiation and Isotopes}
  \textbf{\bibinfo{volume}{63}}, \bibinfo{pages}{235} (\bibinfo{year}{2005}).

\bibitem[{\citenamefont{Sonnabend et~al.}(2011)\citenamefont{Sonnabend,
  Glorius, G{\"o}rres, Kn{\"o}rzer, M{\"u}ller, Sauerwein, Tan, and
  Wiescher}}]{sonnabend2011}
\bibinfo{author}{\bibfnamefont{K.}~\bibnamefont{Sonnabend}},
  \bibinfo{author}{\bibfnamefont{J.}~\bibnamefont{Glorius}},
  \bibinfo{author}{\bibfnamefont{J.}~\bibnamefont{G{\"o}rres}},
  \bibinfo{author}{\bibfnamefont{M.}~\bibnamefont{Kn{\"o}rzer}},
  \bibinfo{author}{\bibfnamefont{S.}~\bibnamefont{M{\"u}ller}},
  \bibinfo{author}{\bibfnamefont{A.}~\bibnamefont{Sauerwein}},
  \bibinfo{author}{\bibfnamefont{W.~P.} \bibnamefont{Tan}}, \bibnamefont{and}
  \bibinfo{author}{\bibfnamefont{M.}~\bibnamefont{Wiescher}},
  \bibinfo{journal}{Journal of Physics: Conference Series}
  \textbf{\bibinfo{volume}{312}}, \bibinfo{pages}{042007}
  (\bibinfo{year}{2011}).

\bibitem[{\citenamefont{Hansen et~al.}(1962)\citenamefont{Hansen, Jopson, Mark,
  and Swift}}]{hansen1962}
\bibinfo{author}{\bibfnamefont{L.~F.} \bibnamefont{Hansen}},
  \bibinfo{author}{\bibfnamefont{R.~C.} \bibnamefont{Jopson}},
  \bibinfo{author}{\bibfnamefont{H.}~\bibnamefont{Mark}}, \bibnamefont{and}
  \bibinfo{author}{\bibfnamefont{C.~D.} \bibnamefont{Swift}},
  \bibinfo{journal}{Nuclear Physics} \textbf{\bibinfo{volume}{30}},
  \bibinfo{pages}{389} (\bibinfo{year}{1962}).

\bibitem[{\citenamefont{{P. Demetriou, C.Grama and S.
  Goriely}}(2002)}]{demetriou2002}
\bibinfo{author}{\bibnamefont{{P. Demetriou, C.Grama and S. Goriely}}},
  \bibinfo{journal}{Nucl. Phys. A} \textbf{\bibinfo{volume}{707}},
  \bibinfo{pages}{253} (\bibinfo{year}{2002}).

\end{thebibliography}

\end{document}